\newtheorem{assum}{Assumption}
\newtheorem{thm}{Theorem}
\newtheorem{cor}{Corollary}
\newtheorem{lem}{Lemma}
\newtheorem{prop}{Proposition}
\newtheorem{defn}{Definition}
\newtheorem{rem}{Remark}

\documentclass[twocolumn]{autart}    

\usepackage{graphicx}          
 
 \usepackage{amsmath}      
 \usepackage{amssymb}
\usepackage{amsbsy}  
 \usepackage[nolist]{acronym}
\usepackage{savesym}
\savesymbol{AND}
\usepackage{algorithm}
\usepackage{algorithmic}
 \usepackage{xcolor}
 \usepackage{enumitem}
 \usepackage{float}
\begin{document}
\pdfminorversion=4

\newcommand{\norm}[1]{\left\Vert#1\right\Vert}
\newcommand{\abs}[1]{\left\vert#1\right\vert}
\newcommand{\set}[1]{\left\{#1\right\}}
\newcommand{\squbra}[1]{\left[#1\right]}
\newcommand{\brackets}[1]{\left(#1\right)}
\newcommand{\Real}{\mathbb R}
\newcommand{\expect}{\mathbb{E}}
\newcommand{\proba}{\mathbb{P}}
\newcommand{\inner}[1]{\left \langle #1 \right \rangle}
\newcommand{\red}[1]{\textcolor{red}{#1}}
\newcommand{\blue}[1]{\textcolor{blue}{#1}}
\newcommand{\qedblack}{\hfill \ensuremath{\blacksquare}}

\begin{frontmatter}

\title{Active Inverse Methods in Stackelberg Games with Bounded Rationality}

\date{\empty}

\author[AMSS]{Jianguo Chen}\ead{chenjianguo@amss.ac.cn}
\author[TJ]{Jinlong Lei}\ead{leijinlong@tongji.edu.cn}
\author[AMSS]{Biqiang Mu}\ead{bqmu@amss.ac.cn}
\author[TJ]{Yiguang Hong}\ead{yghong@iss.ac.cn}
\author[AMSS]{Hongsheng Qi}\ead{qihongsh@amss.ac.cn}

\address[AMSS]{Key Laboratory of Systems and Control, Academy of Mathematics and Systems Science, Chinese Academy of Sciences, Beijing, China; University of Chinese Academy of Sciences, Beijing, China}
\address[TJ]{Department of Control Science and Engineering, Tongji University, Shanghai, China; Shanghai Research Institute for Intelligent Autonomous Systems, Shanghai, China; Shanghai Institute of Intelligent Science and Technology, Tongji University, Shanghai, China; National Key Laboratory of Autonomous Intelligent Unmanned Systems, Shanghai, China; Frontiers Science Center for Intelligent Autonomous Systems, Ministry of Education, Shanghai, China}


\date{\empty}
\maketitle

\begin{abstract}
Inverse game theory is utilized to infer the cost functions of all players based on game outcomes. However, existing inverse game theory methods do not consider the learner as an active participant in the game, which could significantly enhance the learning process. In this paper, we extend inverse game theory to active inverse methods. For Stackelberg games with bounded rationality, the leader, acting as a learner, actively chooses actions to better understand the follower's cost functions. First, we develop a method of active learning by leveraging Fisher information to maximize information gain about the unknown parameters and prove the consistency and asymptotic normality. Additionally, when leaders consider its cost, we develop a method of active inverse game to balance exploration and exploitation, and prove the consistency and asymptotic Stackelberg equilibrium with quadratic cost functions. Finally, we verify the properties of these methods through simulations in the quadratic case and demonstrate that the active inverse game method can achieve Stackelberg equilibrium more quickly through active exploration.
\end{abstract}

\end{frontmatter}

\section{Introduction}
\label{sec:intro}
In recent years, the inverse problem has drawn increasing research attention in different areas such as control systems \cite{yu2021system}, robotics \cite{mombaur2010human}, and machine learning \cite{lian2022inverse}. It aims to infer the system's objective function given the observed optimal behavior, and has been extended from single-agent to multi-agent scenarios.

\begin{figure}[ht]
    \centering
    \includegraphics[width=.5\textwidth]{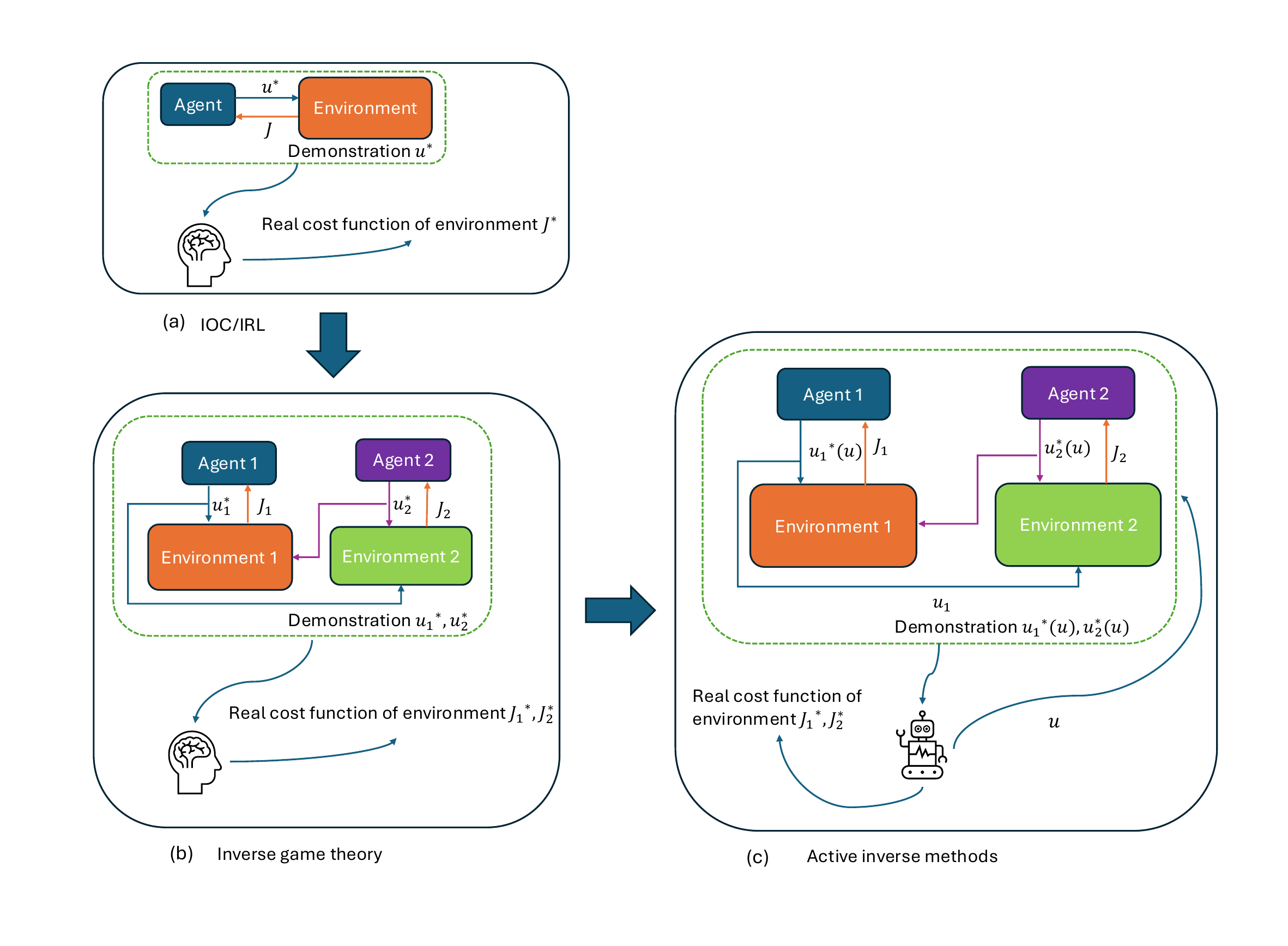}
    \caption{The relationship develops from IOC/IRL to inverse game theory and active inverse methods.}
    \label{fig:inverse}
\end{figure}

In the single-agent case, inverse optimal control (IOC) and inverse reinforcement learning (IRL) are the two most common approaches. Specifically, IOC focuses on deriving an objective function from state/action samples under the assumption of a stable control system \cite{ab2020inverse}. While IRL derives an objective function from expert demonstrations, assuming the expert's behavior is optimal \cite{ab2020inverse}.  Despite structural differences, both methods aim to reconstruct an objective function from observed data, as shown in Fig.~\ref{fig:inverse}~(a). 

In the multi-agent case, inverse game theory is a framework that infers the cost functions of all players by observing the outcomes of the game when the cost functions are unknown \cite{kuleshov2015inverse}, as shown in Fig.~\ref{fig:inverse}~(b). By now, various types of games have been addressed. For example, \cite{molloy2019inverse} aimed to infer parameters of player cost functions given that the state and control trajectories constitute an open-loop Nash equilibrium of a noncooperative differential game, and proposed a method by minimizing the violation of both the costate condition and the Hamiltonian condition. In two-person zero-sum stochastic games, \cite{lin2017multiagent} considered the problem of reconstructing rewards given a minimax bi-policy by solving an equivalent linear program. In addition, \cite{yu2022inverse} inferred the cost function of the players in a matrix game such that a desired joint strategy is a Nash equilibrium based on semidefinite programs and bilevel optimization. However, the aforementioned literature only considers learning the objective function offline after observing all equilibrium data. In contrast, \cite{chen2023online} considers the scenario where equilibria are acquired sequentially, and constructs this parameter identification problem as online optimization. 

Nevertheless, these studies only focus on cases where the source of equilibrium data is given in advance. They do not consider scenarios where the learner, as a player in the game, can influence the outcome. In such cases, the learner can better understand other players' cost functions by actively choosing strategies, as shown in Fig.~\ref{fig:inverse}~(c). Simultaneously, participants in the game have their costs, so one should also consider their costs when actively understanding the opponent. As such, there is a balance between the exploration of active learning and the exploitation of cost reduction. We call the former \textit{active learning} and the latter \textit{active inverse game} to distinguish the two parts, although they are both included in our understanding of active inverse methods. 

In this paper, we specifically consider a Stackelberg game with bounded rationality, where the leader, acting as a learner, can actively choose actions to better learn the true parameters in the follower's cost function. We design an algorithm for active inverse methods in Stackelberg games, focusing on learning and termed \textit{active learning for Stackelberg games}, based on Fisher information. The leader will take the action that maximizes the Fisher information to get more information about the unknown parameters in the follower's cost function.
Additionally, when the leader considers its own interaction costs while learning the follower's cost function, we propose a method of \textit{active inverse game for Stackelberg games} (AIG for Stackelberg games) that achieves the best trade-off between exploration and exploitation like \cite{coggan2004exploration}. The main contributions of this paper are as follows:
\begin{enumerate}
    \item We propose a framework for active inverse games and use the Stackelberg game with bounded
    rationality as an example to illustrate it.
    \item We introduce an active learning approach based on optimal design, where the learner actively maximizes information gained using Fisher information. We prove its consistency and asymptotic normality, and verify its effectiveness through simulations. 
    \item We further consider the scenario with the learner being a player in the game, which also needs to consider its cost that is influenced by other players' strategies. 
    We then propose an active inverse game method for Stackelberg games that effectively balances exploration and exploitation, and prove its consistency as well. 
    \item In particular, for a special case of Stackelberg game with quadratic cost functions, we further provide explicit expressions for all computational functions and prove that the leader's strategy achieves asymptotic Stackelberg equilibrium.
\end{enumerate}
    
The remainder of the paper is organized as follows: Section \ref{sec:formulation} introduces the problem formulation of active inverse methods in Stackelberg games with bounded rationality. Section \ref{sec:learning} presents the active learning algorithm for the problem based on Fisher information. Section \ref{sec:active_cost} derives a method for AIG in Stackelberg games, balancing exploration and exploitation. Section \ref{sec:qua} discusses the active inverse game when players' cost functions are quadratic. The proof of asymptotic consistency is detailed in Section \ref{sec:proof}. Section \ref{sec:simulation} presents simulations and compares the proposed framework with a uniform random strategy and inactive method, demonstrating superior performance.  The paper concludes in Section \ref{sec:conclusion}.

The notation of the paper is shown as follows: $\mathbb{R}$ represents real number set; $\mathbb{R}^n$ represents the n-dimensional vector space; $\mathbb{R}^{n\times m}$ represents the space of $n \times m$-dimensional matrices; If the density function $p(X \mid \theta)$ of a random variable $X$ is influenced by the parameter $\theta$, we abbreviate $\mathbb{E}_{X \sim p(X \mid \theta)}[\cdot]$ and $ \mathrm{Var}_{X \sim p(X \mid \theta)}[\cdot]$ to $\mathbb{E}_{X \mid \theta}[\cdot]$ and $ \mathrm{Var}_{X \mid \theta}[\cdot]$, respectively; Similarly, for $p(X \mid y, \theta)$, we abbreviate $\mathbb{E}_{X \sim p(X \mid y, \theta)}[\cdot]$ and $\mathrm{Var}_{X \sim p(X \mid y, \theta)}[\cdot]$ to $\mathbb{E}_{X \mid y,\theta}[\cdot]$ and $\mathrm{Var}_{X \mid y,\theta}[\cdot]$, respectively; $S(\mathbf{\theta}, \delta) \triangleq \left\{ \mathbf{\theta}' \mid \| \mathbf{\theta}' - \mathbf{\theta} \|_2 \leq \delta \right\}$ denotes a closed neighborhood of radius $\delta$ around the point $\mathbf{\theta}$; $\stackrel{P}{\longrightarrow}$ means convergence in probability; $\stackrel{a.s.}{\longrightarrow}$ means convergence almost everywhere; $\stackrel{L}{\longrightarrow}$ means convergence in distribution; Let \(\mathbf{x} \sim N(\boldsymbol{\mu}, \mathbf{\Sigma})\) denote a multivariate normal distribution with mean vector \(\boldsymbol{\mu}\) and covariance matrix \(\mathbf{\Sigma}\).

\section{Problem formulation} 
\label{sec:formulation}
In this section, we present the problem formulation for active inverse methods in Stackelberg games with bounded rationality.

\subsection{Stackelberg games with bounded rationality}
In a Stackelberg game, the strategic framework outlines two distinct roles: the leader and the follower \cite{von2010market}. Initially, the leader commits to an action, denoted by $\mathbf{u}^L \in \mathbb{R}^n$, selected from a set of feasible actions $\mathbf{U}^L \subseteq \mathbb{R}^n$. Subsequently, the follower observes the leader's chosen action and selects a response action $\mathbf{u}^F \in \mathbb{R}^h$ from its own set of feasible actions $\mathbf{U}^F \subseteq \mathbb{R}^h$. Given this sequential decision-making process, the leader and the follower incur costs denoted by \( J^L(\mathbf{u}^L, \mathbf{u}^F) \) and \( J^F(\mathbf{u}^F, \mathbf{u}^L) \), respectively. The cost functions, \( J^L: \mathbf{U}^L \times \mathbf{U}^F \rightarrow \mathbb{R} \) for the leader and \( J^F: \mathbf{U}^F \times \mathbf{U}^L \rightarrow \mathbb{R} \) for the follower, quantify the respective costs associated with each combination of actions chosen by the leader and the follower.

Given the leader's action \(\mathbf{u}^L\), the follower can select its action \(\mathbf{u}^F\) using various decision-making methods, such as best response or quantal response strategies. In the case of complete rationality, the follower chooses the best response strategy, expressed as:
\begin{equation} \label{eq:best_response}
	\mathbf{u}^F \in \mathop{\arg\min}_{\mathbf{u}^F\in\mathbf{U}^F} J^F(\mathbf{u}^F,\mathbf{u}^L).
\end{equation}
This involves selecting the action from its set \(\mathbf{U}^F\) that minimizes its cost function \(J^F(\mathbf{u}^F, \mathbf{u}^L)\), given the leader's action \(\mathbf{u}^L\) \cite{julien2018stackelberg}. However, in some cases, such as those involving human participants, the follower may not exhibit complete rationality \cite{selten1990bounded,aumann1997rationality}. This study addresses the scenario of bounded rationality. Consistent with prior works \cite{mckelvey1995quantal,goeree2005regular}, we incorporate bounded rationality into the Stackelberg game framework by employing quantal response. Specifically, upon observing the leader's action \(\mathbf{u}^L\), the follower determines its action \(\mathbf{u}^F\) using a probability function:
\begin{equation} \label{pro}
	\begin{split}
	    p(\mathbf{u}^F \mid \mathbf{u}^L) \triangleq \frac{\exp\{-\lambda J^F(\mathbf{u}^F,\mathbf{u}^L)\}}{Z(\mathbf{u}^L)}, \mathbf{u}^F \in \mathbf{U}^F,
	\end{split}
\end{equation} 
where \( Z(\mathbf{u}^L) = \int_{\mathbf{U}^F} \exp\{- \lambda J^F(\mathbf{u}^F,\mathbf{u}^L)\} \, d\mathbf{u}^F \) and the rationality coefficient \(\lambda \geq 0\) is used to modulate the follower's decision-making process \cite{wu2022inverse}. Specifically, when \(\lambda = 0\), the follower completely ignores the leader's influence. As \(\lambda\) approaches positive infinity, the follower's response converges to the best response strategy as \eqref{eq:best_response}.

With knowledge of the cost function and the follower’s rationality coefficient, the leader can strategically determine its optimal deterministic strategy \(\mathbf{u}^{L*}\), defined as the Stackelberg equilibrium strategy. This strategy is given by
\begin{equation}\label{def:Stackelberg_equilibrium_prim}
	\mathbf{u}^{L*} \in \operatorname*{\arg\min}_{\mathbf{u}^L\in\mathbf{U}^L} \mathbb{E}_{\mathbf{u}^F\mid \mathbf{u}^L} \left[J^L(\mathbf{u}^L,\mathbf{u}^F)\right],
\end{equation}
where \(\mathbb{E}_{\mathbf{u}^F\mid\mathbf{u}^L} \left[J^L(\mathbf{u}^L,\mathbf{u}^F)\right]\) denotes the expectation of the cost with respect to the probability density function \(p(\mathbf{u}^F \mid \mathbf{u}^L)\).

\subsection{The problem of active inverse methods in Stackelberg games}
In a Stackelberg game, the leader may encounter a scenario where the follower's cost function is unknown \cite{lauffer2023no,wu2022inverse}. Consequently, the leader lacks the necessary information to compute the Stackelberg equilibrium as defined in \eqref{def:Stackelberg_equilibrium_prim}. To address the uncertainty surrounding the cost function, we adopt a parameterization approach wherein the leader knows the structure but lacks knowledge of the parameter \(\mathbf{\theta}\in \mathbb{R}^m\) within the follower's cost function \(J^F(\mathbf{u}^F,\mathbf{u}^L;\mathbf{\theta})\), with the true parameters denoted as \(\mathbf{\theta}_0\) \cite{lauffer2023no,chen2023online}. The parameter \(\mathbf{\theta}\) is considered to belong to a set \(\mathbf{\Theta} \subset \mathbb{R}^m\). Thus, if the leader assumes the unknown parameter to be \(\mathbf{\theta}\), it anticipates the follower's action selection to be governed by the probability function:
\begin{equation}
\label{density}
	\begin{split}
	    p(\mathbf{u}^F \mid \mathbf{u}^L, \mathbf{\theta}) \triangleq \frac{\exp\{-\lambda J^F(\mathbf{u}^F,\mathbf{u}^L; \mathbf{\theta})\}}{Z(\mathbf{u}^L, \mathbf{\theta})}, \mathbf{u}^F \in \mathbf{U}^F,
	\end{split}
\end{equation} 
where \( Z(\mathbf{u}^L, \mathbf{\theta}) \triangleq \int_{\mathbf{U}^F} \exp\{- \lambda J^F(\mathbf{u}^F,\mathbf{u}^L; \mathbf{\theta})\} \, d\mathbf{u}^F \).


Although in reality, the follower selects its own action based on 
\begin{equation}
\label{true_pro}
	\begin{split}
	    p(\mathbf{u}^F \mid \mathbf{u}^L, \mathbf{\theta}_0) = \frac{\exp\{-\lambda J^F(\mathbf{u}^F,\mathbf{u}^L; \mathbf{\theta}_0)\}}{Z(\mathbf{u}^L, \mathbf{\theta}_0)},\mathbf{u}^F \in \mathbf{U}^F,
	\end{split}
\end{equation} 
following the disclosure of the leader's action $\mathbf{u}^L$. Therefore, the Stackelberg equilibrium strategy in equation~\eqref{def:Stackelberg_equilibrium_prim} is re-expressed by 
\begin{equation}\label{def:Stackelberg_equilibrium}
    \mathbf{u}^{L*} \in \operatorname*{\arg\min}_{\mathbf{u}^L\in\mathbf{U}^L} \mathbb{E}_{\mathbf{u}^F\mid\mathbf{u}^L,\theta_0} \left[J^L(\mathbf{u}^L,\mathbf{u}^F)\right].
\end{equation}

We examine a scenario in which the leader learns the unknown parameters within the follower's cost function through iterative interactions with the follower in a repeated Stackelberg game setting \cite{wu2022inverse,chen2023online}. In the context of repeated Stackelberg games, the leader has the ability to actively employ effective strategies, thereby facilitating a more comprehensive acquisition of the unknown parameters in game-theoretic strategic interaction, with or without considering costs. This paper thus designates this problem as active inverse methods in Stackelberg games.

Specifically, during the \( T \) iterations of the repeated game, the leader actively selects strategies denoted as 
\begin{equation*}
    \mathbf{D}^L(T) \triangleq \{\mathbf{u}^L(t)\}_{t=1}^{T}.
\end{equation*}
This generates a dataset 
\begin{equation*}
    \mathbf{D}(T) \triangleq \{(\mathbf{u}^L(t), \mathbf{u}^F(t))\}_{t=1}^{T},
\end{equation*}
where \( t \) represents the \( t \)-th interaction and \( \mathbf{u}^F(t) \) is selected by the follower according to the probability \eqref{true_pro} under \( \mathbf{u}^L(t) \). Consequently, this iterative process enhances the leader's capability to acquire more accurate knowledge regarding the true parameter \( \mathbf{\theta}_0 \) in game-theoretic strategic interaction, with or without considering costs.


Then we present two commonly employed assumptions regarding the unknown parameter space and cost function \cite{cheng2022single,greenberg1984avoiding,aghassi2006robust,kearns2013graphical}.

\begin{assum}\label{complete}
    The set \(\mathbf{\Theta}\) is compact and convex and $\theta_0$ is an interior point of \(\mathbf{\Theta}\).
\end{assum}

\begin{assum} \label{cost_assu}
    For all \(\mathbf{u}^F \in \mathbf{U}^F\), the function $p(\mathbf{u}^F\mid\mathbf{u}^L,\mathbf{\theta})$  is continuously differentiable with respect to both $\mathbf{\theta} \in \mathbf{\Theta}$ and $\mathbf{u}^L\in \mathbf{U}^L$. Furthermore, it is twice continuously differentiable with respect to $\mathbf{\theta}\in \mathbf{\Theta}$, and the Hessian matrix about $\mathbf{\theta}$ is positive definite and continuous for all $\mathbf{\theta} \in \mathbf{\Theta}$.
\end{assum}

For simplicity, when $\log p(\mathbf{u}^F\mid\mathbf{u}^L,\mathbf{\theta})\neq0$, we define
\begin{equation*}
    \Phi(\mathbf{u}^F; \mathbf{u}^L, \theta) = \log p(\mathbf{u}^F \mid \mathbf{u}^L, \mathbf{\theta}).
\end{equation*}
According to Assumption~\ref{cost_assu}, 
for all \(\mathbf{u}^F \in \mathbf{U}^F\), the functions $\Phi$  is continuously differentiable with respect to both $\mathbf{\theta} \in \mathbf{\Theta}$ and $\mathbf{u}^L\in \mathbf{U}^L$, and let $\dot{\Phi}(\mathbf{u}^F; \mathbf{u}^L, \theta)$ denote the gradient of $\Phi$ with respect to (w.r.t.) $\mathbf{\theta}$. Furthermore, it is twice continuously differentiable w.r.t. $\mathbf{\theta}\in \mathbf{\Theta}$, and the Hessian matrix w.r.t. $\mathbf{\theta}$ of $\Phi$ denoted by $\ddot{\Phi}(\mathbf{u}^F; \mathbf{u}^L, \theta)$ is continuous for all $\mathbf{\theta} \in \mathbf{\Theta}$.

The following assumption is the standard condition for the asymptotic normality of MLE \cite{rao1973linear,hoadley1971asymptotic,van2000asymptotic}.
\begin{assum} \label{interchage}
Given \(\mathbf{u}^L \in \mathbf{U}^L\) and \(\mathbf{\theta} \in \mathbf{\Theta}\), the following properties hold:
    \begin{equation*}
        \frac{\partial }{\partial \theta}\int_{\mathbf{U}^F}p(\mathbf{u}^F \mid \mathbf{u}^L, \mathbf{\theta})d\mathbf{u}^F = \int_{\mathbf{U}^F} \frac{\partial p(\mathbf{u}^F \mid \mathbf{u}^L, \mathbf{\theta})}{\partial \theta} d\mathbf{u}^F,
    \end{equation*}
    and
    \begin{equation*}
        \begin{split}
            \frac{\partial^2 \int_{\mathbf{U}^F}p(\mathbf{u}^F \mid \mathbf{u}^L, \mathbf{\theta})d\mathbf{u}^F }{\partial \theta \partial \theta^{\top}}
        = \int_{\mathbf{U}^F} \frac{\partial^2 p(\mathbf{u}^F \mid \mathbf{u}^L, \mathbf{\theta})}{\partial \theta \partial \theta^{\top}}d\mathbf{u}^F.
        \end{split}
    \end{equation*}
\end{assum}

Assumption~\ref{interchage} ensures that the following relationships hold (see \cite[Theorem 2]{hoadley1971asymptotic}): Given \(\mathbf{u}^L \in \mathbf{U}^L\) and \(\mathbf{\theta} \in \mathbf{\Theta}\),
\begin{equation*}
    \begin{array}{c}
        \mathbb{E}_{\mathbf{u}^F\mid\mathbf{u}^L, \theta} \left[ \dot{\Phi}(\mathbf{u}^F; \mathbf{u}^L, \theta) \right] = 0, \\
        \mathbb{E}_{\mathbf{u}^F\mid\mathbf{u}^L, \theta} \left[ \ddot{\Phi}(\mathbf{u}^F; \mathbf{u}^L, \theta) \right] 
        = - \mathbb{E}_{\mathbf{u}^F\mid\mathbf{u}^L, \theta} \left[ \ddot{\Phi}(\mathbf{u}^F; \mathbf{u}^L, \theta) \right].
    \end{array}
\end{equation*}

\section{Active learning for Stackelberg games} \label{sec:learning}
In this section, we only focus on learning by designing an active learning algorithm for Stackelberg games based on Fisher information.

\subsection{Query strategy based on Fisher information} \label{subsec:query}
The Fisher information quantifies the amount of information that an observable random variable \( X \) carries about an unknown parameter \( \mathbf{\theta}_0 \) on which the probability of \( X \) depends \cite{rohatgi2015introduction,ly2017tutorial}. Let \( p(X \mid \theta_0) \) denote the probability density function of \( X \) conditioned on the value of \( \theta_0 \). Mathematically, the Fisher information $\mathcal{F}(\mathbf{\theta}_0)$ is expressed as (see also\cite[pp. 70–71]{johnson2013statistical})
\begin{equation*}
    \expect_{X \mid  \theta_0}  \left[  \left(\nabla_{\mathbf{\theta}_0} \log p(X \mid \theta_0 ) \right) \left(\nabla_{\mathbf{\theta}_0} \log p(X\mid\theta_0 ) \right)^{\top} \right].
\end{equation*}

In the Stackelberg game described in Section \ref{sec:formulation}, once the leader obtains an estimate \( \hat{\mathbf{\theta}} \) for the unknown parameter \( \mathbf{\theta} \), it aims to devise a query strategy that maximizes the acquisition of informative data. To achieve this goal, leveraging the Fisher information proves to be a promising approach. Accordingly, we give an observation information matrix for the leader in the following definition, which is computed using the estimate \( \hat{\mathbf{\theta}} \) instead of the true parameter \( \mathbf{\theta}_0 \) \cite{jung2021optimal,zeng2022localizability,zeng2022global}.

\begin{defn}
	Given the estimate $\hat{\mathbf{\theta}} \in \mathbf{\Theta}$, the observation information matrix (OIM) $\mathcal{F}(\mathbf{u}^L \mid \hat{\mathbf{\theta}}) \in \mathbb{R}^{m\times m}$ is defined as
	\begin{equation}\label{def-Ful}
		\begin{split}
		    \mathcal{F}(\mathbf{u}^L\mid\hat{\mathbf{\theta}}) \triangleq
        \mathbb{E}_{\mathbf{u}^F\mid \mathbf{u}^L, \hat{\mathbf{\theta}}} \left[ \dot{\Phi}(\mathbf{u}^F; \mathbf{u}^L, \hat{\mathbf{\theta}}) \dot{\Phi}(\mathbf{u}^F; \mathbf{u}^L, \hat{\mathbf{\theta}})^{\top} \right].
		\end{split}
	\end{equation}
\end{defn}

Given the current estimate \( \hat{\mathbf{\theta}} \), we define a query function \( \mathcal{H}(\mathbf{u}^L \mid \hat{\mathbf{\theta}}): \mathbb{R}^n \rightarrow \mathbb{R} \), where the leader seeks a strategy \( \mathbf{u}^L \in \mathbf{U}^L \) that maximizes this function. In the relevant literature, the following representations of \( \mathcal{H} \) are commonly encountered:
\begin{itemize}
\item $\mathcal{H}\brackets{\mathbf{u}^L\mid \hat{\mathbf{\theta}}} = \operatorname{tr}\squbra{\mathcal{F}(\mathbf{u}^L\mid \hat{\mathbf{\theta}})}$, which is known as A-optimality criterion. This criterion aims to maximize the trace of OIM, which is equivalent to minimizing the total parameter variance \cite{lane2020adaptive,he2015fisher,palmin2021fisher}.
	\item $\mathcal{H}\brackets{\mathbf{u}^L\mid \hat{\mathbf{\theta}}} = \brackets{\det\squbra{\mathcal{F}(\mathbf{u}^L\mid \hat{\mathbf{\theta}})}}^{\frac{1}{m}}$, which is known as the D-optimality criterion in experimental design. This criterion seeks to maximize the determinant of OIM, thereby effectively reducing the volume of the joint confidence region for the estimated parameters \cite{xygkis2016fisher,he2015fisher,palmin2021fisher}.
	\item $\mathcal{H}\brackets{\mathbf{u}^L\mid \hat{\mathbf{\theta}}} = \min\operatorname{eig}\squbra{\mathcal{F}(\mathbf{u}^L\mid \hat{\mathbf{\theta}})}$, which is known as the E-optimality criterion. The criterion aims to maximize the minimum eigenvalue of OIM, thereby reducing uncertainties in the most adverse direction within the parameter space \cite{palmin2021fisher,banks2011comparison}.
\end{itemize}

\begin{rem}
\label{rem:query_continuous}
    Under Assumption~\ref{cost_assu}, the continuity in $\mathbf{u}^L$ of these query functions with A-optimality and D-optimality criteria is inherent, because they only perform basic arithmetic operations on elements of $\dot{\Phi}$. For the E-optimality criterion, continuity can be assured through Weyl's theorem on eigenvalue perturbation \cite{weyl1912asymptotische}.
\end{rem}

\subsection{Active learning based on Fisher information}
After the \(t\)-th iteration of repeated games, the leader has gathered data \(\mathbf{D}(T)\) and proceeds to refine its estimate \(\hat{\mathbf{\theta}}(t)\) by maximizing the log-likelihood function \(l_t(\mathbf{\theta}):\mathbb{R}^m\rightarrow\mathbb{R}\) over the parameter space \(\mathbf{\Theta}\) \cite{pan2002maximum}. The log-likelihood function is defined as
\begin{equation}
    \label{mle}
   l_t(\mathbf{\theta}) \triangleq \frac{1}{t} \sum_{k=1}^{t} \Phi(\mathbf{u}^F(k); \mathbf{u}^L(k), \mathbf{\theta}).
\end{equation}

The parameter estimation method above is the maximum likelihood estimation (MLE), and its properties will be analyzed below. For convenience, we temporarily denote \( p(\mathbf{u}^F \mid \mathbf{u}^L, \mathbf{\theta}) \) as \( p \), \( Z(\mathbf{u}^L, \mathbf{\theta}) \) as \( Z \), and \( J^F(\mathbf{u}^F, \mathbf{u}^L; \mathbf{\theta}) \) as \( J^F \), which are defined in \eqref{density}. When $J^F$ is continuously
differentiable with respect to $\mathbf{\theta}$, define
\begin{equation*}
    \begin{split}
        &\text{Cov}\left( \frac{\partial J^F}{\partial \mathbf{\theta}} \right) \\
        =& \mathbb{E}\left[ \left( \frac{\partial J^F}{\partial \mathbf{\theta}} - \mathbb{E}\left[ \frac{\partial J^F}{\partial \mathbf{\theta}} \right] \right) \left( \frac{\partial J^F}{\partial \mathbf{\theta}} - \mathbb{E}\left[ \frac{\partial J^F}{\partial \mathbf{\theta}} \right] \right)^T \right],
    \end{split}
\end{equation*}
which is positive definite.

The following assumption is satisfied when $J^F$ possesses certain properties duo to the the Leibniz Rule in general forms \cite[Theorem 3.2]{lang2013undergraduate}. To simplify the assumptions, we only give the following assumption.
\begin{assum}\label{interchage1}
Given \(\mathbf{u}^L \in \mathbf{U}^L\) and \(\mathbf{\theta} \in \mathbf{\Theta}\), the following properties hold:
    \[
    \frac{\partial}{\partial \mathbf{\theta}} \int_{\mathbf{U}^F} \exp\{-\lambda J^F\} \, d\mathbf{u}^F = \int_{\mathbf{U}^F} \frac{\partial}{\partial \mathbf{\theta}} \exp\{-\lambda J^F\} \, d\mathbf{u}^F,
    \]
    \[
    \frac{\partial \int_{\mathbf{U}^F} \exp\{-\lambda J^F\} \frac{\partial J^F}{\partial \mathbf{\theta}} d\mathbf{u}^F}{\partial \mathbf{\theta}^{\top}} = \int_{\mathbf{U}^F} \frac{\partial \exp\{-\lambda J^F\} \frac{\partial J^F}{\partial \mathbf{\theta}}}{\partial \mathbf{\theta}^{\top}} d\mathbf{u}^F.
    \]
\end{assum}


In certain scenarios, such as when \( J^F \) is linear with respect to \( \mathbf{\theta} \), the log-likelihood function is concave, as shown in the following proposition. 

  \begin{prop}
  \label{thm:likelihood_convex}
      Under Assumptions~\ref{cost_assu} and \ref{interchage1}, the log-likelihood function \eqref{mle} is concave if the term
      \[
      -\lambda \left( \frac{\partial^2 J^F}{\partial \mathbf{\theta} \partial \mathbf{\theta}^T} - \mathbb{E}\left[ \frac{\partial^2 J^F}{\partial \mathbf{\theta} \partial \mathbf{\theta}^T} \right] \right) - \lambda^2 \text{Cov}\left( \frac{\partial J^F}{\partial \mathbf{\theta}} \right)
      \]
      is negative definite.
     In particular, this property holds true when \( J^F \) is linear with respect to \( \mathbf{\theta} \).
  \end{prop}

  \begin{pf}
To prove the concavity of the log-likelihood function, we start by reviewing \eqref{density}. We have 
\[
\log p = -\lambda J^F - \log \left( Z\right).
\]
According to Assumption~\ref{cost_assu}, $\log p$ is twice continuously differentiable with respect to $\mathbf{\theta}$. Next, we will compute the second derivative of \(\log p\) with respect to \(\mathbf{\theta}\).

First, we compute the first derivative of \(\log p\) with respect to \(\mathbf{\theta}\):
\[
\frac{\partial \log p}{\partial \mathbf{\theta}} = -\lambda \frac{\partial J^F}{\partial \mathbf{\theta}} - \frac{1}{Z} \frac{\partial Z}{\partial \mathbf{\theta}}.
\]
Next, we compute the derivative of \( Z\):
\begin{equation*}
    \begin{split}
        \frac{\partial Z}{\partial \mathbf{\theta}} =& \int \frac{\partial}{\partial \mathbf{\theta}} \exp\{-\lambda J^F\} \, d\mathbf{u}^F \\
=& -\lambda \int \exp\{-\lambda J^F\} \frac{\partial J^F}{\partial \mathbf{\theta}} \, d\mathbf{u}^F,
    \end{split}
\end{equation*}
where the first equality holds by Assumption~\ref{interchage1}.
Therefore, we have
\begin{equation*}
        \frac{\partial \log p}{\partial \mathbf{\theta}} = -\lambda \frac{\partial J^F}{\partial \mathbf{\theta}} + \lambda \frac{\int \exp\{-\lambda J^F\} \frac{\partial J^F}{\partial \mathbf{\theta}} \, d\mathbf{u}^F}{Z}.
\end{equation*}
Then we compute the second derivative of \(\log p\):
\begin{equation*}
    \begin{split}
        &\frac{\partial^2 \log p}{\partial \mathbf{\theta} \partial \mathbf{\theta}^T} = -\lambda \frac{\partial^2 J^F}{\partial \mathbf{\theta} \partial \mathbf{\theta}^T} \\
        &+ \lambda \frac{1}{Z^2} \bigg( Z \cdot \frac{\partial}{\partial \theta^T} \int_{\mathbf{U}^F} \exp\{-\lambda J^F\} \frac{\partial J^F}{\partial \mathbf{\theta}} \, d\mathbf{u}^F \\
        &- \frac{\partial Z}{\partial \theta^T} \cdot \int_{\mathbf{U}^F} \exp\{-\lambda J^F\} \frac{\partial J^F}{\partial \mathbf{\theta}} \, d\mathbf{u}^F \bigg) \\
        = & -\lambda \frac{\partial^2 J^F}{\partial \mathbf{\theta} \partial \mathbf{\theta}^T} + \lambda  \expect \left[ \frac{\partial^2 J^F}{\partial \mathbf{\theta} \partial \mathbf{\theta}^T} \right] - \lambda^2  \expect \left[ \frac{\partial J^F}{\partial \theta} \frac{\partial J^F}{\partial \theta^T} \right] \\
        &+ \lambda^2 \expect \left[ \frac{\partial J^F}{\partial \theta}  \right] \expect \left[ \frac{\partial J^F}{\partial \theta^T}  \right] \\
        = & -\lambda \left( \frac{\partial^2 J^F}{\partial \mathbf{\theta} \partial \mathbf{\theta}^T} - \mathbb{E}\left[ \frac{\partial^2 J^F}{\partial \mathbf{\theta} \partial \mathbf{\theta}^T} \right] \right) - \lambda^2 \text{Cov}\left( \frac{\partial J^F}{\partial \mathbf{\theta}} \right).
    \end{split}
\end{equation*}
where the second equality holds by Assumption~\ref{interchage1}.
If \(\frac{\partial^2 \log p}{\partial \mathbf{\theta} \partial \mathbf{\theta}^T}\) is negative definite, the log-likelihood function   \eqref{mle}  is concave.

When \( J^F \) is linear with respect to \(\mathbf{\theta}\), we have \(\frac{\partial^2 J^F}{\partial \mathbf{\theta} \partial \mathbf{\theta}^T} = \mathbf{0}\), then
\[
\frac{\partial^2 \log p}{\partial \mathbf{\theta} \partial \mathbf{\theta}^T} = - \lambda^2 \text{Cov}\left( \frac{\partial J^F}{\partial \mathbf{\theta}} \right).
\]
Thus, \(\frac{\partial^2 \log p}{\partial \mathbf{\theta} \partial \mathbf{\theta}^T}\) is negative definite. Therefore, the log-likelihood function \ is concave.
\qedblack
\end{pf}

Due to the concavity of the function in Proposition~\ref{thm:likelihood_convex} and the convexity of the set $\mathbf{\Theta}$ in Assumption~\ref{complete},  MLE is a convex optimization.

Then we present a common assumption from the literature on parameter identification that ensures the identifiability of \(\mathbf{\theta}_0\) \cite{bohlin1970maximum,pan2002maximum}.

\begin{assum}[Identifiability] \label{iden}
	A necessary and sufficient condition for \(\mathbf{\theta} \neq \mathbf{\theta}_0\) within the parameter space \(\mathbf{\Theta}\) is that \(p(\cdot | \mathbf{u}^L , \mathbf{\theta}) \neq p(\cdot | \mathbf{u}^L , \mathbf{\theta}_0)\) for all \(\mathbf{u}^L \in \mathbf{U}^L\).
 \end{assum}

From the definition of $p(\cdot \mid \mathbf{u}^L , \mathbf{\theta})$, identifiability in Assumption~\ref{iden} means that if $\mathbf{\theta} \neq \mathbf{\theta}_0$,  \(J^F(\cdot, \mathbf{u}^L; \mathbf{\theta}) \neq J^F(\cdot, \mathbf{u}^L; \mathbf{\theta}_0)\) for all \(\mathbf{u}^L \in \mathbf{U}^L\).

When the leader primarily focuses on learning the follower's parameters without incurring any cost, upon updating the estimate \(\hat{\theta}\) using MLE, the leader can achieve this goal by choosing the action
\begin{equation} \label{eq:learning_func}
    \mathbf{u}^{L,\circ } \triangleq \operatorname{\arg\max}_{\mathbf{u}^L \in \mathbf{U}^L} \mathcal{H}(\mathbf{u}^L\mid\hat{\mathbf{\theta}}),
\end{equation}
which maximizes the Fisher information criterion given in Section~\ref{subsec:query}. This ensures the most informative data acquisition regarding the follower's parameters. Consequently, the active learning algorithm based on Fisher information is outlined in Algorithm \ref{alg:fisher}.

 \begin{algorithm}[htp]
	\begin{algorithmic}[1]
	\REQUIRE $T, \lambda, \hat{\mathbf{\theta}}(0)$;
		\FOR{$t=1, 2, \ldots, T$}
		\STATE The leader chooses its action \(\mathbf{u}^{L,\circ}(t)\) under estimate \(\hat{\mathbf{\theta}}(t-1)\) by using \eqref{eq:learning_func}.
		\STATE  The follower observes the leader's action  \(\mathbf{u}^{L, \circ}(t)\), and updates its action \(\mathbf{u}^F(t)\) according to \eqref{true_pro}.
        \STATE  The leader obverses the response of the follower \(\mathbf{u}^F(t)\).
        \STATE The leader gathers the data \(\mathbf{D}(t)=\left\{\right(\mathbf{u}^{L,\circ }(k), \mathbf{u}^F(k)\left)\right\}_{k=1}^{t}\).
        \STATE The leader then gives an estimate \(\hat{\mathbf{\theta}}(t)\)  of the follower's true parameter $\theta_0$ by MLE, where \(\hat{\mathbf{\theta}}(t)\) maximizes \(l_t\) in \eqref{mle} under data $\mathbf{D}(t)$.
		\ENDFOR
		\ENSURE \(\hat{\mathbf{\theta}}(T)\)
	\end{algorithmic}
	\caption{Active learning algorithm}
	\label{alg:fisher}
\end{algorithm}

\subsection{Consistency and asymptotic normality of Alg. \ref{alg:fisher}} 
\label{sec:theory_alg1}

We will give the consistency and asymptotic normality of Algorithm~\ref{alg:fisher}. First of all, some boundedness assumptions are given to ensure stability and tractability in models involving bounded rationality.
\begin{assum}\label{assum:comprehensive}
    Given $\mathbf{u}^L \in \mathbf{U}^L$ and $\mathbf{\theta} \in \mathbf{\Theta}$, 
    \begin{enumerate}
        \item \(\operatorname{Var}_{\mathbf{u}^F \mid \mathbf{u}^L, \mathbf{\theta}} [J^F(\mathbf{u}^F, \mathbf{u}^L ; \mathbf{\theta})]\) is bounded.
        \item \(\mathbb{E}_{\mathbf{u}^F \mid \mathbf{u}^L, \mathbf{\theta}} \| \dot{\Phi}(\mathbf{u}^F; \mathbf{u}^L, \mathbf{\theta}) \|^3\) is finite.
        \item There exist $\epsilon > 0$ and random variables $B_{ij}(\mathbf{u}^F)$ such that 
      $  \sup \{|\ddot{\Phi}_{i,j}(\mathbf{u}^F; \mathbf{u}^L, \theta)| : || (\theta^\top,\mathbf{u}^{L\top})  - (\theta_0^\top,\mathbf{u}_0^{L\top})|| \leq \epsilon \} \leq B_{ij}(\mathbf{u}^F)$
    and $\expect|B_{ij}(\mathbf{u}^F)|^{1+\delta}$ is bounded.
        \item \(\mathbb{E}_{\mathbf{u}^F \mid \mathbf{u}^L, \mathbf{\theta}} \left[ \max_{\mathbf{\theta} \in \mathbf{\Theta}} | (\dot{\Phi}^\top \dot{\Phi})_{i,j} | \right]\) is finite for all \(\mathbf{u}^F \in \mathbf{U}^F\).
    \end{enumerate}
\end{assum}

These assumptions ensure stability and tractability by bounding variances, gradients, and higher-order terms. Bounded variance limits stochastic fluctuations of the cost function, while the bounded third moment of gradients controls extreme values. Assumptions~\ref{assum:comprehensive}. (3) and (4) further ensure the feasibility of parameterized systems. Relevant boundedness assumptions are widely used in the literature to ensure stability and tractability in models with bounded rationality \cite{fudenberg1998theory,camerer2006does}.

The following theorem demonstrates that the estimates obtained by Algorithm \ref{alg:fisher} converge to the true parameters \(\mathbf{\theta}_0\) in probability. Before giving the theorem, we review the notations \(\mathbf{D}(T) = \{ \mathbf{u}^{L,\circ }(t) ,\mathbf{u}^F(t))\}_{t=1}^{T}\), \(\mathbf{D}^L(T) = \{\mathbf{u}^{L,\circ}(t)\}_{t=1}^{T}\), 
and define
\begin{equation} \label{lll}
	\begin{split}
		l_{T}\brackets{\mathbf{\theta} | \mathbf{D}(T)} \triangleq & \frac{1}{T} \sum_{t=1}^{T}\log p(\mathbf{u}^F(t) \mid \mathbf{u}^{L}(t), \mathbf{\theta}), \\
 L_T(\mathbf{\theta} | \mathbf{D}^L(T) ) \triangleq & \frac{1}{T} \sum_{t=1}^{T} \expect_{\mathbf{u}^F \mid \mathbf{u}^L(t),\mathbf{\theta}_0} \squbra{\log p(\mathbf{u}^F \mid \mathbf{u}^{L}(t),\mathbf{\theta})}.
	\end{split}
\end{equation}
\begin{thm}[Consistency of Alg. \ref{alg:fisher}] \label{consistency_alg1}
	 Under Assumptions \ref{complete}, \ref{iden} and \ref{assum:comprehensive}, the estimates \(\hat{\mathbf{\theta}}(T)\) obtained from Algorithm \ref{alg:fisher} exhibit consistency in probability. Namely, for any \(\epsilon > 0\),
\begin{align}
    \lim_{T\rightarrow\infty}\mathbb{P}\left\{\|\hat{\mathbf{\theta}}(T) - \mathbf{\theta}_0\| \geq \epsilon \right\} = 0.
\end{align}
which is denoted as \(\hat{\mathbf{\theta}}(T) \stackrel{P}{\longrightarrow} \mathbf{\theta}_0\).
\end{thm}

\noindent \textbf{Proof Outline.}
The proof of this theorem is detailed in Section \ref{sec:proof}, which consists of the following two steps:

\textit{Step 1:}
Using the law of large numbers to prove that
\begin{equation*}
        \lim_{T\rightarrow\infty}\proba \left\{ \left|l_{T}(\hat{\mathbf{\theta}}(T) | \mathbf{D}(T) ) - L_{T}(\mathbf{\theta}_0 | \mathbf{D}^L(T) )\right| \geq \epsilon \right\} = 0.
\end{equation*}

\textit{Step 2:} Using proof by contradiction to get the final conclusion.
\qedblack

In order to analyze  the active action $\mathbf{u}^{L,\circ}$, we define
\begin{equation}\label{eq:max_uL}
    \mathbf{u}^{L\circ\circ} \in \operatorname{\arg\max}_{\mathbf{u}^L \in \mathbf{U}^L} \mathcal{H}(\mathbf{u}^L\mid\mathbf{\theta}_0).
\end{equation}
The following theorem gives the convergence of the active action obtained by Algorithm~\ref{alg:fisher}.
\begin{thm}\label{var}
Under Assumptions \ref{complete}, \ref{cost_assu}, \ref{iden} and \ref{assum:comprehensive}, if the maximum $\mathbf{u}^{L\circ\circ}$ is unique, we have
\[ \mathbf{u}^{L\circ}(T) \overset{P}{\longrightarrow} \mathbf{u}^{L\circ\circ}, \quad \text{as} \quad T\rightarrow\infty. \]
\end{thm}

\begin{pf}
Firstly, we have
\begin{equation*}
	\begin{split}
		0 \leq& \mathcal{H}\brackets{\mathbf{u}^{L\circ\circ} \mid \mathbf{\theta_0}} - \mathcal{H}\brackets{\mathbf{u}^{L\circ}(T) \mid \mathbf{\theta_0}} \\
		=& \mathcal{H}\brackets{\mathbf{u}^{L\circ\circ} \mid \mathbf{\theta_0}} - \mathcal{H}\brackets{\mathbf{u}^{L\circ\circ} \mid \hat{\theta}(T)} \\
  &+ \mathcal{H}\brackets{\mathbf{u}^{L\circ\circ} \mid \hat{\theta}(T)}  - \mathcal{H}\brackets{\mathbf{u}^{L\circ}(T) \mid \hat{\theta}(T)} \\
  &+ \mathcal{H}\brackets{\mathbf{u}^{L\circ}(T) \mid \hat{\theta}(T)} - \mathcal{H}\brackets{\mathbf{u}^{L\circ}(T) \mid \mathbf{\theta_0}} \\
		\leq & \mathcal{H}(\mathbf{u}^{L\circ\circ} \mid \mathbf{\theta_0}) - \mathcal{H}\brackets{\mathbf{u}^{L\circ\circ} \mid \hat{\theta}(T)}  \\
  &+ \mathcal{H}\brackets{\mathbf{u}^{L\circ}(T) \mid \hat{\theta}(T)} - \mathcal{H}\brackets{\mathbf{u}^{L\circ}(T) \mid \mathbf{\theta_0}},
	\end{split}
\end{equation*}
where the last inequality holds because 
\begin{equation*}
    \mathcal{H}\brackets{\mathbf{u}^{L\circ\circ} \mid \hat{\theta}(T)} - \mathcal{H}\brackets{\mathbf{u}^{L\circ}(T) \mid \hat{\theta}(T)} \leq 0,
\end{equation*}
according to equation \eqref{eq:learning_func}  of Algorithm \ref{alg:fisher}.
As a consequence, we derive
\begin{equation}
\label{eq:ineq_event}
    \begin{split}
      &\abs{\mathcal{H}\brackets{\mathbf{u}^{L\circ\circ} \mid \mathbf{\theta_0}} - \mathcal{H}\brackets{\mathbf{u}^{L\circ}(T) \mid \mathbf{\theta_0}}} \\
      \leq& \abs{\mathcal{H}(\mathbf{u}^{L\circ\circ} \mid \mathbf{\theta_0}) - \mathcal{H}\brackets{\mathbf{u}^{L\circ\circ} \mid \hat{\theta}(T)}} \\
      &+ \abs{\mathcal{H}\brackets{\mathbf{u}^{L\circ}(T) \mid \hat{\theta}(T)} - \mathcal{H}\brackets{\mathbf{u}^{L\circ}(T) \mid \mathbf{\theta_0}}}.
    \end{split}
\end{equation}


Due to Assumption \ref{cost_assu} and Assumption~\ref{assum:comprehensive}.(4) , by using Lebesgue's dominated convergence theorem \cite[Theorem 1.34]{Rudin1968RealAC}  we can establish that the Fisher information \(\mathcal{F}(\mathbf{u}^L \mid \mathbf{\theta})\)  defined by \eqref{def-Ful} is  continuous with respect to \(\mathbf{\theta}\) for any \(\mathbf{u}^L \in \mathbf{U}^L\). Additionally, \(\mathcal{H}(\mathbf{u}^L \mid \mathbf{\theta})\) exhibits continuity at \(\theta_0\) for all \(\mathbf{u}^L \in \mathbf{U}^L\) from Remark \ref{rem:query_continuous}.
Therefore, for any \( \epsilon > 0 \), there exists \( \delta > 0 \) such that for all \( \theta \in S(\mathbf{\theta}_0, \delta) \cap \mathbf{\Theta} \), it holds that \( |\mathcal{H}(\mathbf{u}^{L} \mid \mathbf{\theta_0}) - \mathcal{H}(\mathbf{u}^{L} \mid \theta)| < \epsilon / 2 \) for all \( \mathbf{u}^L \in \mathbf{U}^L \). Particularly, this implies
\begin{equation*}
    \begin{array}{l}
         |\mathcal{H}(\mathbf{u}^{L\circ\circ} \mid \mathbf{\theta_0}) - \mathcal{H}(\mathbf{u}^{L\circ\circ} \mid \hat{\theta}(T))| < \frac{\epsilon}{2},  \\
         |\mathcal{H}(\mathbf{u}^{L\circ}(T), \hat{\theta}(T)) - \mathcal{H}(\mathbf{u}^{L\circ}(T), \mathbf{\theta_0})| < \frac{\epsilon}{2}, 
    \end{array}
\end{equation*}
for all \( \hat{\theta}(T)) \in S(\mathbf{\theta}_0, \delta) \cap \mathbf{\Theta} \).

Moreover, in conjunction with \eqref{eq:ineq_event}, it can be demonstrated that
\begin{equation*}
\begin{split}
  & \proba \set{\hat{\mathbf{\theta}}(T) \in S(\mathbf{\theta}_0, \delta) \cap \mathbf{\Theta}} \\
  \leq & \proba \set{\abs{\mathcal{H}\brackets{\mathbf{u}^{L\circ\circ} \mid \mathbf{\theta_0}} - \mathcal{H}\brackets{\mathbf{u}^{L\circ}(T) \mid \mathbf{\theta_0}}} < \epsilon} \\
  \leq & 1.
\end{split}
\end{equation*}
Because  $\hat{\mathbf{\theta}}(T) \overset{P}{\longrightarrow} \theta_0 $ as $T \rightarrow \infty$ from Theorem \ref{consistency_alg1}, we obtain
\[
\lim_{T\rightarrow\infty} \proba \set{\abs{\mathcal{H}\brackets{\mathbf{u}^{L\circ\circ} \mid \mathbf{\theta_0}} - \mathcal{H}\brackets{\mathbf{u}^{L\circ}(T) \mid \mathbf{\theta_0}}} < \epsilon} = 1.
\]
This implies that
\[ \mathcal{H}\brackets{\mathbf{u}^{L\circ}(T) \mid \mathbf{\theta_0}} \overset{P}{\longrightarrow} \mathcal{H}\brackets{\mathbf{u}^{L\circ\circ} \mid \mathbf{\theta_0}} \quad \text{as} \quad T \rightarrow \infty. \]
Consequently, we deduce
\[ \mathbf{u}^{L\circ}(T) \overset{P}{\longrightarrow} \mathbf{u}^{L\circ\circ}, \quad \text{as} \quad T\rightarrow\infty, \]
from Remark \ref{rem:query_continuous} and the uniqueness of \(\mathbf{u}^{L\circ\circ}\). The proof is completed. \qedblack
\end{pf}

To elucidate additional properties of Algorithm \ref{alg:fisher}, we introduce the following notation: 
\begin{equation}\label{eq:F_T_def}
    \begin{array}{l}
        \mathcal{F}_T(\mathbf{\theta}) \triangleq \frac{1}{T}\sum_{t=1}^{T}\mathcal{F}(\mathbf{u}^{L\circ}(t) \mid \mathbf{\theta}).
    \end{array} 
\end{equation}
Now, we present the following corollary concerning \( \mathcal{F}_T \), which is crucial for proving Theorem \ref{as_no}.
\begin{cor}\label{cor:F2Fmax}
Under Assumptions \ref{complete}, \ref{cost_assu}, \ref{iden} and \ref{assum:comprehensive}, if the maximum $\mathbf{u}^{L\circ\circ}$ is unique, we have that for any $\theta \in \Theta$
    \begin{equation}
        \mathcal{F}_T(\mathbf{\theta}) \overset{P}{\longrightarrow} \mathcal{F}(\mathbf{u}^{L\circ\circ} \mid \mathbf{\theta}).
    \end{equation}
\end{cor}

\begin{pf} 
Because $\mathcal{F}$ is continuous w.r.t. $\mathbf{u}^F$ from Remark~\ref{rem:query_continuous} under Assumption~\ref{cost_assu} and
\begin{equation*}
    \mathbf{u}^{L\circ}(t) \overset{P}{\longrightarrow} \mathbf{u}^{L\circ\circ}, \quad \text{as} \quad t\rightarrow\infty,
\end{equation*}
from Theorem~\ref{var}, we obtain
\begin{equation*}
    \mathcal{F}(\mathbf{u}^{L\circ}(t) \mid \mathbf{\theta})\overset{P}{\longrightarrow} \mathcal{F}(\mathbf{u}^{L\circ\circ} \mid \mathbf{\theta}).
\end{equation*}
Furthermore, it follows that \(\mathcal{F}_T(\mathbf{\theta}) \overset{P}{\longrightarrow} \mathcal{F}(\mathbf{u}^{L\circ\circ} \mid \mathbf{\theta})\) as \(T \rightarrow \infty\). Thus, the conclusion follows. \qedblack
\end{pf}

$\mathcal{F}(\mathbf{u}^{L\circ\circ} \mid \mathbf{\theta})$ is related to the asymptotic variance of the estimate, and the details will be given in the subsequent theorem. But before that, let us first introduce the following lemma.
 \begin{lem}[Theorem 2 in \cite{hoadley1971asymptotic}]
 \label{lem:asymptotic}
Let $Y_1, Y_2, \ldots$ be a sequence of independent random variables. Assume that $Y_t$ has density $p_t(y_t \mid \theta_0)$. And the MLE of the true parameter value $\theta_0$ is denoted by $\hat{\theta}_n \in \mathbf{\Theta}$, which maximizing the log-likelihood
\[ l_T(\theta) = \frac{1}{T} \sum_{t=1}^n \Phi_t(y_t, \theta), \]
like equation \eqref{mle}, where $\Phi_{t}(y_t, \theta)=\log p_t(y_t \mid \theta)$.
If the following conditions \textit{1)} to \textit{9)} are satisfied, then
\[
\sqrt{T}(\hat{\theta}(T) - \theta_0) \xrightarrow{L} N(0, \Gamma^{-1}(\theta_0)).
\]
\begin{enumerate}[label=\arabic*).]
    \item $\theta_0$ is an interior point of $\mathbf{\Theta}$.
    \item $\hat{\theta}(T) \xrightarrow{P} \theta_0$.
    \item $\dot{\Phi}_{t}(y_t,\theta) = \frac{\partial \log p_t(y_t \mid \theta)}{\partial \theta}$ and $\ddot{\Phi}_{t}(y_t,\theta) = \frac{\partial^2 \log p_t(y_t \mid \theta)}{\partial \theta \partial \theta^{\top}}$ exist.
    \item  For all $t$, $\ddot{\Phi}_{t}(y_t,\theta)$ is a continuous function of $\theta$ and is a measurable function of $Y_t$.
    \item $\expect[\dot{\Phi}_{t}(y_t,\theta) \mid \theta] = 0 \quad \text{for} \quad t = 1, 2, \cdots$.
    \item $\Gamma_t(\theta) \triangleq \expect[\dot{\Phi}_{t}(y_t,\theta) \dot{\Phi}_{t}(y_t,\theta)^{\top} | \theta] = -\expect[\ddot{\Phi}_{t}(y_t,\theta) | \theta],$ for $t = 1, 2, \cdots$.
    \item For some $\delta > 0$ and all \( \theta \in S(\mathbf{\theta}_0, \delta) \cap \mathbf{\Theta} \), if $T^{-1}\sum_{k}\left[ - \ddot{\Phi}_{t,ij}(y_t,\theta) \right] - T^{-1}\sum_{k}\Gamma_{t,ij}(\theta) \overset{P}{\to} 0$, then $n^{-1}\sum_{t}\left[ - \ddot{\Phi}_{t,ij}(y_t,\theta) \right] \overset{P}{\to} \Gamma_{ij}(\theta)$ holds, where $\Gamma(\theta)$ is positive definite.
    \item $\expect\norm{\frac{\partial \log p_t(y \mid \theta)}{\partial \theta}}^3$ is bounded.
    \item There exist $\epsilon > 0$ and random variables $B_{t,ij}(Y_t)$ such that $\sup \{|\frac{\partial^2 \log p_t(y \mid \theta)}{\partial \theta_i \partial \theta_j^{\top}}| : ||\theta - \theta_0|| \leq \epsilon \} \leq B_{t,ij}(Y_t)$ and $\expect|B_{t,ij}(Y_t)|^{1+\delta}$ is bounded.
\end{enumerate}
 \end{lem}
 
 \begin{pf}
Among these conditions, only \textit{1)} and \textit{7)} are different from those in the literature \cite{hoadley1971asymptotic}. However, these do not affect the proof process because of the following two points. 
\begin{enumerate}
    \item The original condition \textit{1)} in \cite{hoadley1971asymptotic} states that `$\Theta$ is an open subset of $\mathbb{R}^m$', implying the existence of $\eta > 0$ such that $S(\theta_0, \eta) \subset \mathbf{\Theta}$. Thus, instead of requiring `$\theta_0$ is an interior point of $\mathbf{\Theta}$', this condition suffices. 
    \item The original condition \textit{7)} in \cite{hoadley1971asymptotic} is only used to ensure that for some $\delta > 0$ and all \( \theta \in S(\mathbf{\theta}_0, \delta) \cap \mathbf{\Theta} \), if $T^{-1}\sum_{t}\left[ - \ddot{\Phi}_{t,ij}(y_t,\theta) \right] - T^{-1}\sum_{t}\Gamma_{t,ij}(\theta) \overset{P}{\to} 0$, then $T^{-1}\sum_{t}\left[ - \ddot{\Phi}_{t,ij}(y_t,\theta) \right] \overset{P}{\to} \Gamma_{ij}(\theta)$ holds, where $\Gamma(\theta)$ is positive definite. Therefore, we directly use the conclusion to replace the original condition \textit{7)}.
\end{enumerate}
In addition to these, the proofs in \cite{hoadley1971asymptotic} remain applicable with this substitution. \qedblack
\end{pf}

By employing Corollary~\ref{cor:F2Fmax}  and Lemma~\ref{lem:asymptotic}, we establish the following theorem, affirming that the algorithm establishes the asymptotic normality of the estimate and achieves optimal estimate under a certain metric.

\begin{thm}[Asymptotic normality of Alg. \ref{alg:fisher}] \label{as_no}
Under Assumptions \ref{complete}, \ref{cost_assu}, \ref{interchage}, \ref{iden} and \ref{assum:comprehensive}, if the maximum $\mathbf{u}^{L\circ\circ}$ is unique, and $\mathcal{F}(\mathbf{u}^{L\circ\circ} \mid \mathbf{\theta})$ is positive definite, then the estimates \(\hat{\theta}(T)\) obtained from Algorithm  \ref{alg:fisher} converge to a normal distribution as \(T \rightarrow \infty\):
\[
\sqrt{T}\left(\hat{\theta}(T) - \theta_0\right) \stackrel{L}{\longrightarrow} N\left(0, \mathcal{F}(\mathbf{u}^{L\circ\circ}, \mathbf{\theta}_0)^{-1}\right),
\]
where \(\mathbf{u}^{L\circ\circ}\) is defined in \eqref{eq:max_uL}.
\end{thm}

\begin{pf} 
Our proof process primarily relies on the conclusions from Lemma \ref{lem:asymptotic}, and the conditions are verified as follows.

\textit{(A)}. Condition~\textit{1)} in Lemma~\ref{lem:asymptotic} is satisfied by Assumption~\ref{complete}.

\textit{(B)}. Condition~\textit{2)} in Lemma~\ref{lem:asymptotic} is satisfied by Theorem~\ref{consistency_alg1}.

\textit{(C)}. Under Assumption~\ref{cost_assu}, conditions \textit{3)} and \textit{4)} in Lemma~\ref{lem:asymptotic} are satisfied.

\textit{(D)}. Condition~\textit{5)} and Condition~\textit{6)} in Lemma~\ref{lem:asymptotic} are satisfied by Assumption~\ref{interchage}.

\textit{(E)}. For some $\delta > 0$ and all \( \theta \in S(\mathbf{\theta}_0, \delta) \cap \mathbf{\Theta} \), assume 
\begin{equation}\label{eq:condition_original}
    T^{-1}\sum_{t}\left[ - \ddot{\Phi}_{t,ij}(\mathbf{u}^F(t),\theta) \right] - \mathcal{F}_T(\mathbf{\theta}) \overset{P}{\to} 0,
\end{equation}
where $\mathcal{F}_T(\mathbf{\theta})$ is defined in \eqref{eq:F_T_def}. From Corollary~\ref{cor:F2Fmax},
\begin{equation}\label{eq:F2Fmax}
    \mathcal{F}_T(\mathbf{\theta}) \overset{P}{\longrightarrow} \mathcal{F}(\mathbf{u}^{L\circ\circ} \mid \mathbf{\theta}),
\end{equation}
where $\mathcal{F}(\mathbf{u}^{L\circ\circ} \mid \mathbf{\theta})$ is positive definite. Therefore, combining equations~\ref{eq:condition_original}~and~\ref{eq:F2Fmax}, we have
\begin{equation*}
    T^{-1}\sum_{t}\left[ - \ddot{\Phi}_{t,ij}(\mathbf{u}^F(t),\theta) \right] \overset{P}{\to} \mathcal{F}(\mathbf{u}^{L\circ\circ} \mid \mathbf{\theta}).
\end{equation*}
Thus, condition~\textit{7)} in Lemma~\ref{lem:asymptotic} is satisfied.

\textit{(F)}. It follows straight forwardly that conditions~\textit{8)}~and~\textit{9)} hold by Assumption~\ref{assum:comprehensive}.

Therefore, from Lemma~\ref{lem:asymptotic}, the estimates \(\hat{\theta}(T)\) obtained from Alg. \ref{alg:fisher}  converge to a normal distribution as \(T \rightarrow \infty\):
\[
\sqrt{T}\left(\hat{\theta}(T) - \theta_0\right) \stackrel{L}{\longrightarrow} N\left(0, \mathcal{F}(\mathbf{u}^{L\circ\circ}, \mathbf{\theta}_0)^{-1}\right),
\]
where \(\mathbf{u}^{L\circ\circ}\) is defined in \eqref{eq:max_uL}. Thus, the conclusion follows. \qedblack
\end{pf}

From the definition of $\mathbf{u}^{L\circ\circ}$ in \eqref{eq:max_uL}, it can be concluded that the algorithm achieves a certain optimal optimality to ensure that the asymptotic variance is as small as possible in the sense of MLE.

\section{Active inverse game for Stackelberg games}\label{sec:active_cost}
In this section, we focus on the scenario where the leader also considers its cost while actively learning the unknown parameter in the follower's cost function.

\subsection{Balance between exploitation and exploration} \label{sec:balance}
In Section \ref{sec:learning}, we have investigated a scenario where the leader merely focuses on learning the follower's cost function while disregarding its own costs. However, the follower also acts as a player within the game, necessitating consideration of its own costs influenced by the strategies of other players. This interdependence between strategies complicates both exploration (attempting to discover more information about the unknown parameters by selecting actions to increase the amount of Fisher information) and exploitation (leveraging existing knowledge to achieve the locally optimal solutions) tasks. Therefore, we now explore strategies for leader to effectively balance exploration and exploitation to achieve optimal trade-offs \cite{coggan2004exploration}.

\begin{rem}
    The difference from reinforcement learning (RL): The exploration-exploitation trade-off also exists in RL \cite{kaelbling1996reinforcement,yogeswaran2012reinforcement}. However, RL typically involves a single agent learning from an environment without explicit interaction with opponents. Different from  RL agents that explore and exploit based on environmental feedback, active inverse methods incorporate game-theoretic feedback into the learning process, as the strategies of players influence each other. Moreover, active inverse methods aim to achieve equilibrium efficiently, unlike RL, which focuses solely on optimizing rewards without considering strategic equilibrium between multiple players.
\end{rem}

If the leader's estimate of the unknown parameter is $\hat{\mathbf{\theta}}$, then the estimated expected cost of the action $\mathbf{u}^L$ is
\begin{equation} \label{eq:expect_cost}
    \mathcal{J}(\mathbf{u}^L \mid \hat{\mathbf{\theta}}) \triangleq \mathbb{E}_{\mathbf{u}^F \mid \mathbf{u}^L,\hat{\mathbf{\theta}}} \left[J^L(\mathbf{u}^L,\mathbf{u}^F; \hat{\mathbf{\theta}})\right],
\end{equation}
which is based entirely on current information. 
 
Next, we propose a query strategy that effectively balances exploitation and exploration. At the \(t\)-th iteration of repeated games, the leader selects action \(\mathbf{u}^{L\diamond}(t)\) by solving the following optimization problem:
\begin{equation} \label{query2}
	\begin{split}
	  \min_{\mathbf{u}^L \in \mathbf{U}^L} &\underbrace{ \mathcal{J}(\mathbf{u}^L \mid \hat{\mathbf{\theta}}(t)) }_{\text{exploitation}} - \underbrace{\rho_t\mathcal{H}(\mathbf{u}^L\mid\hat{\mathbf{\theta}}(t))}_{\text{exploration}},\\
	\end{split}
\end{equation}
where \(\rho_t\) is a carefully chosen coefficient that balances exploitation and exploration, as detailed later. In \eqref{query2}, the \textit{exploitation} term represents the expected cost of the leader's action based on current knowledge, aiming to achieve locally optimal results known so far; the \textit{exploration} term utilizes the Fisher information associated with the leader's action, as discussed in Section~\ref{sec:learning}, to discover more new information regarding the follower. Because $\mathcal{H}$ is always greater than 0, we set $\rho_t \rightarrow 0$ as $t \rightarrow \infty$ to reduce the focus on learning the follower's information after a certain level of learning.

Except for the query strategy, the process is the same as Algorithm~\ref{consistency_alg1} in Section~\ref{sec:active_cost}.
The algorithm of active inverse game for Stackelberg games, which balances cost and learning, is summarized in Algorithm~\ref{alg2}.
\begin{algorithm}[ht]
	\begin{algorithmic}[1]
	\REQUIRE $T, \lambda, \hat{\mathbf{\theta}}(0)$;
		\FOR{$t=1, 2, \ldots, T$}
        \STATE The leader computes the coefficient $\rho_t$.
		\STATE The leader chooses its action \(\mathbf{u}^{L\diamond }(t)\) under estimate \(\hat{\mathbf{\theta}}(t-1)\) by using \eqref{query2}.
		\STATE The leader broadcasts \(\mathbf{u}^{L\diamond }(t)\) to the follower.
        \STATE The follower obverses the response of the follower \(\mathbf{u}^F(t)\).
        \STATE The leader gathers the data \(\mathbf{D}(t)=\left\{\right(\mathbf{u}^{L\diamond }(k), \mathbf{u}^F(k)\left)\right\}_{k=1}^{t}\).
        \STATE The leader then finds the \(\hat{\mathbf{\theta}}(t)\) by MLE, where \(\hat{\mathbf{\theta}}(t)\) maximizes \(l_t\) in \eqref{mle} under data $\mathbf{D}(t)$.
		\ENDFOR
		\ENSURE \(\hat{\mathbf{\theta}}(T)\)
	\end{algorithmic}
	\caption{Active inverse game algorithm}
	\label{alg2}
\end{algorithm}

\begin{rem}
Specifically, we give an example of the coefficient design which has good experimental results.
Set
\begin{equation*}
    \rho_t \triangleq \mu_t \sigma\left(\alpha\left(\|\hat{\mathbf{\theta}}(t)-\hat{\mathbf{\theta}}(t-1)\|_2 - \eta\right)\right),
\end{equation*}
where \(\mu_t\) is a non-negative decreasing sequence such that \(\mu_t \rightarrow 0\) as \(t \rightarrow \infty\), and \(\sigma(x) = \frac{1}{1 + e^{-x}}\) denotes the standard sigmoid function.
Here, \(\alpha > 0\) is a large enough constant, and \(\eta > 0\) is a small threshold. Thus, 
\begin{enumerate} 
\item If \(\|\hat{\mathbf{\theta}}(t)-\hat{\mathbf{\theta}}(t-1)\|_2 > \eta\),
\[ \sigma\left(\alpha\left(\|\hat{\mathbf{\theta}}(t)-\hat{\mathbf{\theta}}(t-1)\|_2 - \eta\right)\right)\approx1 \]
duo to the large enough constant $\alpha$. Then \(\rho_t \approx \mu_t\);
\item Similarly, if \(\|\hat{\mathbf{\theta}}(t)-\hat{\mathbf{\theta}}(t-1)\|_2 < \eta\), then \(\rho_t \approx 0\);
\item As \(t\) becomes sufficiently large, \(\rho_t \approx 0\).
\end{enumerate}
This formulation is guided by the following rationale:
\begin{enumerate}
    \item In the early stages of interaction, when information about the parameters is limited, the estimation of expected costs may be inaccurate, necessitating an emphasis on exploration efficiency.
    \item The convergence rate of the variance in MLE parameters is known to be \(O(\frac{1}{t})\). As the number of interactions increases, parameter learning variance improves. When the difference between parameter estimates is small, i.e., \(\|\hat{\mathbf{\theta}}(t)-\hat{\mathbf{\theta}}(t-1)\|_2 < \eta\), the strategy should focus more on exploitation to minimize estimated costs effectively.
\end{enumerate}
\end{rem}

The following theorem states that the estimates obtained by Algorithm \ref{alg2} converge in probability to the true parameters \(\mathbf{\theta}_0\).

\begin{thm}[Consistency of Alg. \ref{alg2}] \label{consistency_alg2}
Under Assumptions \ref{complete}, \ref{iden} and \ref{assum:comprehensive}, the estimates \(\hat{\mathbf{\theta}}(T)\) obtained from Algorithm~\ref{alg2} exhibit consistency in probability, denoted as \(\hat{\mathbf{\theta}}(T) \stackrel{P}{\longrightarrow} \mathbf{\theta}_0\).
\end{thm}

\begin{pf}
    The proof process is similar to Theorem~\ref{consistency_alg1}, and the detailed process will be given in Section \ref{sec:proof}. \qedblack
\end{pf}
	
 
\subsection{Convergence to Stackelberg equilibrium} \label{sec:qua}
In this part, we will introduce the consistency of actions under the quadratic cost function. 
The cost functions of the leader and follower are specified as: \begin{equation}
\label{eq:quad}
    \begin{split}
        J^L(\mathbf{u}^L,\mathbf{u}^F) =&  \frac{1}{2}\mathbf{u}^{L\top}\mathbf{Q}^L\mathbf{u}^L + \mathbf{u}^{F\top}\mathbf{R}_1^L\mathbf{u}^L \\
        &+ \frac{1}{2}\mathbf{u}^{F\top}\mathbf{R}_2^L\mathbf{u}^F,\\
J^F(\mathbf{u}^F,\mathbf{u}^L,\mathbf{\theta}_0) =&  \frac{1}{2}\mathbf{u}^{F\top}\mathbf{Q}(\mathbf{\theta}_0)\mathbf{u}^F + \mathbf{u}^{L\top}\mathbf{R}_1^F\mathbf{u}^F \\
&+ \frac{1}{2}\mathbf{u}^{L\top}\mathbf{R}_2^F\mathbf{u}^L,
    \end{split}
\end{equation}
where $\mathbf{Q}^L$, $\mathbf{Q}(\mathbf{\theta}_0)$ are symmetric positive definite matrices, and $\mathbf{R}_2^L$, $\mathbf{R}_2^F$ are symmetric matrices. This game framework has many potential applications, such as modeling public goods games in \cite{gorelik2021stackelberg}. The matrix $\mathbf{Q}(\mathbf{\theta}_0)$ depends on the unknown parameter $\theta$, with $\theta_0$ representing its true value. 

Therefore, for the quadratic game  shown in \eqref{eq:quad}, the probability function \eqref{density} can be refined to
\begin{equation}
 \begin{split}
     &p(\mathbf{u}^F \mid \mathbf{u}^L, \mathbf{\theta}) = \frac{1}{Z(\mathbf{u}^L, \mathbf{\theta})} \exp\{-\lambda J^F(\mathbf{u}^F,\mathbf{u}^L, \mathbf{\theta})\} \\
     =& (2\pi )^{-h/2}\det({\boldsymbol {\Sigma }(\mathbf{\theta})})^{-1/2}\,\\
     &\cdot \exp \set{-{\frac {1}{2}}(\mathbf{u}^F -\mathbf{\mu}(\mathbf{\theta}))^{\mathrm {T} }{\boldsymbol {\Sigma }}(\mathbf{\theta})^{-1}(\mathbf{u}^F -\mathbf{\mu}(\mathbf{\theta}))} \\
 \end{split}
\end{equation}
where $\mathbf{\mu}(\mathbf{\theta}) \triangleq -\mathbf{Q}^{-1}(\mathbf{\theta})\mathbf{R}_1^F \mathbf{u}^L$ and $\mathbf{\Sigma}(\mathbf{\theta}) \triangleq \frac{1}{\lambda} \mathbf{Q}(\mathbf{\theta})^{-1}$. Thus, $\mathbf{u}^F$ follows a multivariate normal distribution with mean $\mathbf{\mu}(\mathbf{\theta})$ and covariance $\mathbf{\Sigma}(\mathbf{\theta})$ in the quadratic games shown in \eqref{eq:quad}. Furthermore, Assumption \ref{iden} is simplified to: \textit{The sufficient and necessary condition for $\mathbf{\theta} \neq \mathbf{\theta}_0$ within the parameter space $\mathbf{\Theta}$ is that $Q(\mathbf{\theta}) \neq Q(\mathbf{\theta}_0)$ for all $\mathbf{u}^L \in \mathbf{U}^L$}.

First of all, in order to analyze the properties of the function $\mathcal{J}$ in \eqref{eq:expect_cost},  
we define
\begin{equation}
\label{eq:c}
 \begin{split}
     \mathbf{C}(\theta) \triangleq& \mathbf{Q}^L + 2\mathbf{R}_1^{F\textsf{T}} \mathbf{Q}(\theta)^{-1}\mathbf{R}_2^L\mathbf{Q}(\theta)^{-1} \mathbf{R}_1^F \\
     &- \mathbf{R}_1^{F\textsf{T}} \mathbf{Q}(\theta)^{-1}\mathbf{R}_1^L  - \mathbf{R}_1^{L\textsf{T}} \mathbf{Q}(\theta)^{-1}\mathbf{R}_1^F.
 \end{split}
\end{equation}

Next, we give an assumption about the continuity of function $\mathbf{Q}(\mathbf{\theta})$ and positive definiteness of $\mathbf{C}(\mathbf{\theta}_0)$.

\begin{assum}
\label{c_positive}
\(\mathbf{Q}(\theta)\) is a continuous function of \(\theta\), and \(\mathbf{C}(\theta_0)\) is positive definite.
\end{assum}

Assumption~\ref{c_positive} can ensure that the expected cost with respect to $\mathbf{u}^L$ is strongly convex, which will be rigorously given in the following lemma.

\begin{lem}
\label{thm:convex_cost}
Under Assumption \ref{c_positive}, there exists \(\eta > 0\) such that for all \(\hat{\theta} \in S(\theta_0, \eta)\), \(\mathcal{J}(\mathbf{u}^L \mid \hat{\mathbf{\theta}})\) is strongly convex with respect to \(\mathbf{u}^F \in \mathbf{U}^F\).
\end{lem}

\begin{pf}
From Assumption~\ref{c_positive}, the matrix \(\mathbf{Q}(\theta)\) is continuous. Then the matrix expression \(\mathbf{C}(\theta)\), which involves \(\mathbf{Q}(\theta)\) and its inverse, is also continuous with respect to \(\theta\), since matrix operations like addition, multiplication, and inversion preserve continuity.
Therefore, there exists \(\eta > 0\) such that for all \(\hat{\theta} \in S(\theta_0, \eta)\), the matrix \(\mathbf{C}(\hat{\theta})\) is positive definite.

Note that to compute the expectation of the quadratic form \(\mathbb{E}[\mathbf{x}^T \mathbf{Q} \mathbf{x}]\) for a multivariate normal random vector \(\mathbf{x} \sim N(\boldsymbol{\mu}, \mathbf{\Sigma})\), we can use the formula:

\[
\mathbb{E}[\mathbf{x}^T \mathbf{Q} \mathbf{x}] = \text{tr}(\mathbf{Q} \mathbf{\Sigma}) + \boldsymbol{\mu}^T \mathbf{Q} \boldsymbol{\mu},
\]

where \(\text{tr}(\cdot)\) denotes the trace of a matrix, which is the sum of its diagonal elements.
Since \( \mathbf{u}^F \) follows a normal distribution with mean $\mathbf{\mu}(\mathbf{\theta}) \triangleq -\mathbf{Q}^{-1}(\mathbf{\theta})\mathbf{R}_1^F \mathbf{u}^L$ and covariance matrix $\mathbf{\Sigma}(\mathbf{\theta}) \triangleq \frac{1}{\lambda} \mathbf{Q}(\mathbf{\theta})^{-1}$, 
, we can conclude from \eqref{eq:quad} that
    \begin{equation}
    \begin{split}
        &\mathcal{J}(\mathbf{u}^L \mid \hat{\mathbf{\theta}}) = \mathbb{E}_{\mathbf{u}^F \mid \mathbf{u}^L,\hat{\theta}} \left[J^L\left(\mathbf{u}^F, \mathbf{u}^L, \hat{\theta}\right)\right] \\
        =&\frac{1}{2}\mathbf{u}^{L\top}\mathbf{Q}^L\mathbf{u}^L + \mathbf{\mu}\brackets{\hat{\theta}}^{\top}\mathbf{R}_1^L\mathbf{u}^L + \operatorname{tr} \left[\mathbf{R}_2^L \mathbf{\Sigma}\brackets{\hat{\theta}} \right]\\
        &+\mathbf{\mu}\brackets{\hat{\mathbf{\theta}}} ^{T}\mathbf{R}_2^L \mathbf{\mu}\brackets{\hat{\mathbf{\theta}}} \\
        =&\frac{1}{2}\mathbf{u}^{L\top}\brackets{\mathbf{Q}^L+2\mathbf{R}_1^{F\textsf{T}} \mathbf{Q}^{-1}\mathbf{R}_2^L\mathbf{Q}^{-1} \mathbf{R}_1^F}\mathbf{u}^L \\
        &- \mathbf{u}^{L\top}\mathbf{R}_1^{F\textsf{T}} \mathbf{Q}^{-1}\mathbf{R}_1^L\mathbf{u}^L +  \operatorname{tr} \left[\mathbf{R}_2^L \mathbf{\Sigma}\brackets{\hat{\theta}} \right] \\
        =& \frac{1}{2} \mathbf{u}^{L\top} \mathbf{C}(\hat{\theta}) \mathbf{u}^L + \operatorname{tr} \left[\mathbf{R}_2^L \mathbf{\Sigma}\brackets{\hat{\theta}} \right].
    \end{split}
\end{equation}
Thus, we conclude that \(\mathcal{J}(\mathbf{u}^L \mid \hat{\mathbf{\theta}})\) is convex with respect to \(\mathbf{u}^F\). \qedblack
\end{pf}

With the property of function $\mathcal{J}$ given in Lemma~\ref{thm:convex_cost}, we can obtain the asymptotic property about Stackelberg equilibrium in Algorithm~\ref{alg2}.

\begin{thm}
\label{as_no2}
Under Assumptions \ref{complete}, \ref{iden}, \ref{assum:comprehensive} and \ref{c_positive}, the actions \(\mathbf{u}^{L\diamond}(t)\) generated by Algorithm \ref{alg2} converge to the Stackelberg equilibrium. Namely, as \(T \rightarrow \infty\),
\[ \mathbf{u}^{L\diamond}(T) \stackrel{P}{\longrightarrow} \mathbf{u}^{L*}, \]
where \(\mathbf{u}^{L*}\) represents the Stackelberg equilibrium of the leader as defined in \eqref{def:Stackelberg_equilibrium}.
\end{thm}

\begin{pf}
From Theorem \ref{consistency_alg2}, it follows that \(\hat{\theta}(T) \stackrel{P}{\longrightarrow} \theta_0\) as \(T \rightarrow \infty\). Therefore, there exists a sufficiently large \(T_0\) such that for \(T > T_0\), \(\rho_T\) is sufficiently small, and with probability \(1 - \varepsilon_T\) (\(\varepsilon_T \rightarrow 0\)), \(\hat{\mathbf{\theta}}(T) \in S(\theta_0, \eta)\). Furthermore, according to Lemma~\ref{thm:convex_cost},
\[
\mathcal{J}(\mathbf{u}^L \mid \hat{\mathbf{\theta}}(T))  - \rho_T \mathcal{H}(\mathbf{u}^L \mid \hat{\mathbf{\theta}}(T)),
\]
is strongly convex with respect to \(\mathbf{u}^L\), ensuring a unique minimum \(\mathbf{u}^{L\diamond}(T)\).
Therefore, 
\begin{equation*}
    \begin{split}
        \nabla_{\mathbf{u}^L} \mathcal{J}(\mathbf{u}^{L\diamond}(T) \mid \hat{\mathbf{\theta}}(T)) 
        - \rho_T \nabla_{\mathbf{u}^L}\mathcal{H}(\mathbf{u}^{L\diamond}(T) \mid \hat{\mathbf{\theta}}(T)) = 0.
    \end{split}
\end{equation*}
Given \(\rho_T \rightarrow 0\), it follows that
\[
\lim_{T \rightarrow \infty} \nabla_{\mathbf{u}^L} \mathcal{J}(\mathbf{u}^{L\diamond}(T) \mid \hat{\mathbf{\theta}}(T)) = 0,
\]
implying \(\mathbf{u}^{L\diamond}(T) \rightarrow \mathbf{u}^{L*}\).
This convergence demonstrates that \(\mathbf{u}^{L\diamond}(T)\) asymptotically approaches the Stackelberg equilibrium action \(\mathbf{u}^{L*}\) of the leader. \qedblack
\end{pf}



Theorem~\ref{as_no2} establishes the asymptotic properties of Algorithm~\ref{alg2} in infinite time. Additionally, Section~\ref{sec:balance} has discussed its properties in finite time, which will be further verified in the simulation part. Note that while the empirical results suggest that the algorithm performs well in finite time, a theoretical analysis of its finite-time behavior is not yet provided and remains an open question for future investigation.

\section{Proof of consistency}
\label{sec:proof}

In this section, we present the proof of consistency for  Algorithm  \ref{alg:fisher} and \ref{alg2}. The proofs of Theorems \ref{consistency_alg1} and \ref{consistency_alg2} are similar, and thus, we demonstrate them uniformly. First, we review the notations \(\mathbf{D}(T) = \{(\mathbf{u}^{L}(t),\mathbf{u}^F(t))\}_{t=1}^{T}\) and \(\mathbf{D}^L(T) = \{\mathbf{u}^{L}(t)\}_{t=1}^{T}\) which are obtained by either Algorithm \ref{alg:fisher} or Algorithm \ref{alg2}.

To establish consistency, we first introduce the following lemma.

\begin{lem} \label{lemma1}
Under the identifiability of $\theta_0$ in Assumption \ref{iden}, for all $\mathbf{\theta} \neq \mathbf{\theta}_0$ in $\mathbf{\Theta}$, there exists $\epsilon > 0$ such that
\[ L_{T}(\mathbf{\theta}_0 \mid \mathbf{D}^L(T) ) > L_{T}(\mathbf{\theta} \mid \mathbf{D}^L(T) ) + \epsilon, \]
where $L_{T}$ is defined in \eqref{lll}.
\end{lem}
\begin{pf}
For all $\mathbf{u}^L \in \mathbf{U}^L$, consider the difference
\begin{equation}\label{diff}
    \begin{split}
    &\expect_{\mathbf{u}^F \mid \mathbf{u}^L, \mathbf{\theta}_0} \left[\log\frac{ p(\mathbf{u}^F \mid \mathbf{u}^L,\mathbf{\theta})}{p(\mathbf{u}^F \mid \mathbf{u}^L,\mathbf{\theta}_0)}\right] \\
     =&  \expect_{\mathbf{u}^F \mid \mathbf{u}^L, \mathbf{\theta}_0} \squbra{\log p(\mathbf{u}^F \mid \mathbf{u}^L,\mathbf{\theta})} \\
     &- \expect_{\mathbf{u}^F \mid \mathbf{u}^L, \mathbf{\theta}_0} \squbra{\log p(\mathbf{u}^F \mid \mathbf{u}^L,\mathbf{\theta_0}}).
    \end{split}
\end{equation}
Under Assumption \ref{iden}, we have \( \frac{p(\mathbf{u}^F \mid \mathbf{u}^L,\mathbf{\theta})}{p(\mathbf{u}^F \mid \mathbf{u}^L,\mathbf{\theta}_0)} \neq 1 \) for all $\mathbf{\theta} \neq \mathbf{\theta}_0$ in $\mathbf{\Theta}$. Since $\log t < t - 1$ for any $t \neq 1$, we deduce that for all $\mathbf{u}^L \in \mathbf{U}^L$,
	\begin{equation*}
		\begin{split}
			&\expect_{\mathbf{u}^F \mid \mathbf{u}^L, \mathbf{\theta}_0} \left[\log\frac{ p(\mathbf{u}^F | \mathbf{u}^L,\mathbf{\theta})}{p(\mathbf{u}^F | \mathbf{u}^L,\mathbf{\theta}_0)}\right] \\
   < & \expect_{\mathbf{u}^F \mid \mathbf{u}^L, \mathbf{\theta}_0} \left[\frac{ p(\mathbf{u}^F | \mathbf{u}^L,\mathbf{\theta})}{p(\mathbf{u}^F | \mathbf{u}^L,\mathbf{\theta}_0)}-1\right] \\
			=& \int_{\mathbf{U}^F} \left(\frac{ p(\mathbf{u}^F | \mathbf{u}^L,\mathbf{\theta})}{p(\mathbf{u}^F | \mathbf{u}^L,\mathbf{\theta}_0)}-1\right) p(\mathbf{u}^F | \mathbf{u}^L,\mathbf{\theta}_0) d\mathbf{u}^F\\
			=& \int_{\mathbf{U}^F}  p(\mathbf{u}^F | \mathbf{u}^L,\mathbf{\theta}) d\mathbf{u}^F - \int_{\mathbf{U}^F}  p(\mathbf{u}^F | \mathbf{u}^L,\mathbf{\theta}_0) d\mathbf{u}^F \\
   =& 1-1 = 0.
		\end{split}
	\end{equation*}
Let \( \epsilon = - \inf_{\mathbf{\theta}\in\mathbf{\Theta}} \min_{\mathbf{u^L}\in\mathbf{U}^L } \expect_{\mathbf{u}^F \mid \mathbf{u}^L, \mathbf{\theta}_0} \left[\log\frac{ p(\mathbf{u}^F | \mathbf{u}^L,\mathbf{\theta})}{p(\mathbf{u}^F | \mathbf{u}^L,\mathbf{\theta}_0)}\right]\). Therefore, we can derive \(\epsilon>0\) from Assumption \ref{complete}. This together with \eqref{diff}  demonstrates that 
	\begin{equation*}
		\begin{split}
			&L_{T}(\mathbf{\theta} | \mathbf{D}^L(T) ) - L_{T}(\mathbf{\theta}_0 | \mathbf{D}^L(T) ) \\
			= &\frac{1}{T} \sum_{t=1}^{T} \bigg[\expect_{\mathbf{u}^F \mid \mathbf{u}^L, \mathbf{\theta}_0}[\log p(\mathbf{u}^F | \mathbf{u}^L(t),\mathbf{\theta})]\\
   &-\expect_{\mathbf{u}^F \mid \mathbf{u}^L, \mathbf{\theta}_0}[\log p(\mathbf{u}^F \mid \mathbf{u}^L(t),\mathbf{\theta}_0)] \bigg] \\
			\leq & - \epsilon.
		\end{split}
	\end{equation*}
	The proof is completed. \qedblack
\end{pf}

The following lemma gives an intermediate result, which is one of the sufficient conditions for the Kolmogorov strong law of large numbers to hold \cite{jiang2010large}.

\begin{lem}\label{kol_condition}
 Under Assumption~\ref{assum:comprehensive}, we have that for all \(\mathbf{\theta} \in \mathbf{\Theta}\),
    \[\sum_{t=1}^{\infty} \frac{1}{t^2} \mathrm{Var}_{\mathbf{u}^F \mid \mathbf{u}^L, \mathbf{\theta}_0} \squbra{\log p(\mathbf{u}^F | \mathbf{u}^L(t), \mathbf{\theta})} < \infty. \]
\end{lem}
\begin{pf}
By reviewing the definition of $p(\mathbf{u}^F \mid \mathbf{u}^L(t), \mathbf{\theta})$ in \eqref{density}, we have
    \begin{equation*}
        \begin{split}
            &\sum_{t=1}^{\infty} \frac{1}{t^2} \mathrm{Var}_{\mathbf{u}^F \mid \mathbf{u}^L, \mathbf{\theta}_0} \squbra{\log p(\mathbf{u}^F \mid \mathbf{u}^L(t), \mathbf{\theta})}
            \\
            = & \sum_{t=1}^{\infty} \frac{1}{t^2} \mathrm{Var}_{\mathbf{u}^F \mid \mathbf{u}^L, \mathbf{\theta}_0} \squbra{- \lambda J^F(\mathbf{u}^F \mid   \mathbf{u}^L(t), \mathbf{\theta}))-\log(Z(\mathbf{u}^L,\theta))}
            \\ = 
            &\sum_{t=1}^{\infty} \frac{\lambda^2}{t^2} \mathrm{Var}_{\mathbf{u}^F \mid \mathbf{u}^L, \mathbf{\theta}_0} \squbra{J^F(\mathbf{u}^F \mid   \mathbf{u}^L(t), \mathbf{\theta}))}.
        \end{split}
    \end{equation*}
    Because $\mathrm{Var}_{\mathbf{u}^F \mid \mathbf{u}^L, \mathbf{\theta}_0} \squbra{J^F(\mathbf{u}^F  \mathbf{u}^L(t), \mathbf{\theta}))}$ is bounded from Assumpton~\ref{assum:comprehensive}.(1). Thus, we get
    \[
    \sum_{t=1}^{\infty} \frac{1}{t^2} \mathrm{Var}_{\mathbf{u}^F \mid \mathbf{u}^L, \mathbf{\theta}_0} \squbra{\log p(\mathbf{u}^F | \mathbf{u}^L(t), \mathbf{\theta})} < \infty. \] The proof is completed. \qedblack
\end{pf}


\subsection*{Proof of Theorems \ref{consistency_alg1} and \ref{consistency_alg2}:}
Since the estimate $\hat{\theta}(T)$ is the maximizer of $l_T(\mathbf{\theta} \mid \mathbf{D}(T) )$ obtained through maximum likelihood estimation,
\begin{equation} \label{123}
    \begin{split}
        &l_{T}\brackets{\mathbf{\theta}_0 | \mathbf{D}(T)} - L_{T}(\mathbf{\theta}_0 | \mathbf{D}^L(T) ) \\
        \leq& l_{T}(\hat{\mathbf{\theta}}(T) | \mathbf{D}(T) ) - L_{T}(\mathbf{\theta}_0 | \mathbf{D}^L(T) ).
    \end{split}
\end{equation}
Additionally, from Lemma \ref{lemma1}, we know that $\mathbf{\theta}_0$ is the maximizer of $L_T(\mathbf{\theta} \mid \mathbf{D}^L(T))$. Therefore, 
\begin{equation} \label{456}
	\begin{split}
	    & l_{T}(\hat{\mathbf{\theta}}(T) | \mathbf{D}(T) ) - L_{T}(\mathbf{\theta}_0 | \mathbf{D}^L(T) ) \\
     \leq& l_{T}(\hat{\mathbf{\theta}}(T) | \mathbf{D}(T) ) - L_{T}(\hat{\mathbf{\theta}}(T) | \mathbf{D}^L(T) ).
	\end{split}
\end{equation}
Taking into account both (\ref{123}) and (\ref{456}), we can deduce
\begin{equation*}
	\begin{split}
		 \big| l_{T}(\hat{\mathbf{\theta}}&(T) | \mathbf{D}(T) ) - L_{T}(\mathbf{\theta}_0 | \mathbf{D}^L(T) )\big| \\
\leq  \max \bigg\{ & \left|l_{T}(\hat{\mathbf{\theta}}(T) | \mathbf{D}(T) ) - L_{T}(\hat{\mathbf{\theta}}(T) | \mathbf{D}^L(T) )\right|, \\
 & \left|l_{T}(\mathbf{\theta}_0 | \mathbf{D}(T) ) - L_{n}(\mathbf{\theta}_0 | \mathbf{D}^L(T) )\right|  \bigg\}.
	\end{split}
\end{equation*}

Moreover, for any \(\epsilon > 0\), it can be derived that the event
\[
\left\{ \left|l_{T}(\hat{\mathbf{\theta}}(T) | \mathbf{D}(T) ) - L_{T}(\mathbf{\theta}_0 | \mathbf{D}^L(T) )\right| \geq \epsilon \right\}
\] 
is a subset of the event 
\begin{equation*}
    \begin{split}
        &\left\{ \left|l_{T}(\hat{\mathbf{\theta}}(T) | \mathbf{D}(T) ) - L_{T}(\hat{\mathbf{\theta}}(T) | \mathbf{D}^L(T) )\right| \geq \epsilon \right\} \\
        \cdot \bigcup& \left\{ \left|l_{T}(\mathbf{\theta}_0 | \mathbf{D}(T) ) - L_{T}(\mathbf{\theta}_0 | \mathbf{D}^L(T) )\right| \geq \epsilon \right\}.
    \end{split}
\end{equation*}
Therefore, we  derive
\begin{equation} \label{twoleq}
    \begin{split}
        &\proba \left\{ \left|l_{T}(\hat{\mathbf{\theta}}(T) | \mathbf{D}(T) ) - L_{T}(\mathbf{\theta}_0 | \mathbf{D}^L(T) )\right| \geq \epsilon \right\} \\
        \leq& \proba\left\{ \left|l_{T}(\hat{\mathbf{\theta}}(T) | \mathbf{D}(T) ) - L_{T}(\hat{\mathbf{\theta}}(T) | \mathbf{D}^L(T) )\right| \geq \epsilon \right\} \\
        &+ \proba\left\{ \left|l_{T}(\mathbf{\theta}_0 | \mathbf{D}(T) ) - L_{T}(\mathbf{\theta}_0 | \mathbf{D}^L(T) )\right| \geq \epsilon \right\}.
    \end{split}
\end{equation}

Let’s review the Kolmogorov strong law of large numbers \cite[Theorem 2.3.10]{jiang2010large}: Let
$X_i, i > 1$, be independent random variables such that $\mathbb{E} X_i= \mu_i$ and $\operatorname{Var}(X_i) = \sigma_i^2$ exist
for every $i \geq 1$, and let $\bar{\mu}_n = n^{-1} \sum_{i=1}^{n}\mu_i$. Then if $\sum_{k\geq1} k^{-2}\sigma_i^2 < \infty$, then $\bar{X}_n - \bar{\mu}_n$
 $\stackrel{P}{\longrightarrow} 0$. Therefore, by reviewing the definitions of $l_{T}(\mathbf{\theta} | \mathbf{D}(T) )$ and $L_{T}(\mathbf{\theta} | \mathbf{D}^L(T) )$ in \eqref{lll} and the result of  Lemma \ref{kol_condition}, we can apply Kolmogorov strong law of large numbers to directly get that for any \( \mathbf{\theta} \in \mathbf{\Theta}\), as $T\rightarrow\infty$,
\begin{equation} \label{1234}
    l_{T}(\mathbf{\theta} | \mathbf{D}(T) ) - L_{T}(\mathbf{\theta} | \mathbf{D}^L(T) ) \stackrel{P}{\longrightarrow}  0.
\end{equation}
	Thus,
\begin{equation} \label{twolim}
    \begin{split}
        \lim_{T\rightarrow\infty} &\proba\left\{ \left|l_{T}(\hat{\mathbf{\theta}}(T) | \mathbf{D}(T) ) - L_{T}(\hat{\mathbf{\theta}}(T) | \mathbf{D}^L(T) )\right| \geq \epsilon \right\} = 0, \\
        \lim_{T\rightarrow\infty} &\proba\left\{ \left|l_{T}(\mathbf{\theta}_0 | \mathbf{D}(T) ) - L_{T}(\mathbf{\theta}_0 | \mathbf{D}^L(T) )\right| \geq \epsilon \right\} = 0.
    \end{split}
\end{equation}

Combining (\ref{twoleq}) and (\ref{twolim}), we conclude that 
\[
\lim_{T\rightarrow\infty} \proba\left\{ \left|l_{T}(\hat{\mathbf{\theta}}(T) | \mathbf{D}(T) ) - L_{T}(\mathbf{\theta}_0 | \mathbf{D}^L(T) )\right| \geq \epsilon \right\} = 0,
\]
\textit{i.e.}, as $T\rightarrow\infty$, $l_{T}(\hat{\mathbf{\theta}}(T) | \mathbf{D}(T) ) - L_{T}(\mathbf{\theta}_0 | \mathbf{D}^L(T) ) \stackrel{P}{\longrightarrow}  0$. According to \eqref{1234} and the convergence property of line combination of two random variables by probability, we infer that as $T\rightarrow\infty$,
\begin{equation} \label{contra}
    L_T (\hat{\mathbf{\theta}}(T) | \mathbf{D}^L(T) ) - L_{T}(\mathbf{\theta}_0 | \mathbf{D}^L(T) ) \stackrel{P}{\longrightarrow}  0.
\end{equation}

Then we use proof by contradiction to establish that \(\hat{\theta}(T) \overset{P}{\longrightarrow} \theta_0\) as \(T \rightarrow \infty\). Assume that \(\hat{\theta}(T) \overset{P}{\longrightarrow} \theta_0\) does not hold. In other words, there exist \(\epsilon_0 > 0\) and \(\delta_0 > 0\) such that, for any \(T > 0\), there exists \(t_0 > T\) with \(\proba \left\{ \|\hat{\theta}(t_0) - \theta_0 \| \geq \epsilon_0 \right\} > \delta_0\).
From Lemma \ref{lemma1}, we know that for any \(\theta \notin S(\mathbf{\theta}_0, \epsilon_0) \), there exists \(\epsilon > 0\) such that \(L_{t_0}(\mathbf{\theta} | \mathbf{D}^L(t_0) ) - L_{t_0}(\mathbf{\theta}_0 | \mathbf{D}^L(t_0) ) < - \epsilon\). Therefore,
\begin{equation*}
    \begin{split}
        &\proba\left\{ \left|L_{t_0}(\hat{\theta}(t_0) | \mathbf{D}^L(t_0) ) - L_{t_0}(\mathbf{\theta}_0 | \mathbf{D}^L(t_0) )\right| \geq  \epsilon \right\} \\
        \geq& \proba \left\{\hat{\theta}(t_0) \notin S(\mathbf{\theta}_0, \epsilon_0) \right\} \\
        =&  \proba \left\{ \|\hat{\theta}(t_0) - \theta_0 \| \geq \epsilon_0 \right\} \\
        >& \delta_0.
    \end{split}
\end{equation*}
This contradicts with (\ref{contra}). Thus, the proof is completed. \qedblack

\section{Numerical Simulations}
\label{sec:simulation}
Consider a Stackelberg game involving a leader who also acts as a learner and a follower, where their cost functions are quadratic as described in \eqref{eq:quad}. 

We set
\begin{equation*}
    \begin{split}
    & \mathbf{Q}^L=\left[\begin{matrix} 41 & 2 \\ 2 & 8 \end{matrix} \right],
      \mathbf{R}_1^L=\left[\begin{matrix} 12 & 42 \\ 13 & 1 \end{matrix} \right],
      \mathbf{R}_2^L=\left[\begin{matrix} 400 & 34 \\ 34 & 4 \end{matrix} \right], \\
    & \mathbf{Q}(\mathbf{\theta}_0) = \left[\begin{matrix} \mathbf{\theta}_{10} & \mathbf{\theta}_{20} \\ \mathbf{\theta}_{20} & \mathbf{\theta}_{30} \end{matrix} \right]
      \text{and} \ \mathbf{R}_1^F=\left[\begin{matrix} 16 & 8 \\ 9 & 31 \end{matrix} \right],
    \end{split}
\end{equation*}
where \(\mathbf{Q}(\mathbf{\theta}_0)\) is a symmetric positive definite matrix and $\mathbf{\theta}_0= [\mathbf{\theta}_{10},\mathbf{\theta}_{20},\mathbf{\theta}_{30}]$ is the unknown parameter to be estimated.
The term \(\mathbf{R}_2^F\) in \eqref{eq:quad} is disregarded as it does not influence the follower's decision-making. 

We set the truth value of the parameter \(\mathbf{\theta} = [\mathbf{\theta}_1,\mathbf{\theta}_2,\mathbf{\theta}_3]\) to \(\mathbf{\theta}_0 = [20,10,30]\) and its value space to \(\mathbf{\Theta}=\{\mathbf{\theta} | \mathbf{\theta}_1 \geq \kappa, \mathbf{\theta}_1\mathbf{\theta}_3-\mathbf{\theta}_2^2 \geq \kappa \}\), where $\kappa$ represents a small positive number to ensure that $\mathbf{Q}(\mathbf{\theta})$ is a positive definite matrix and $\mathbf{\Theta}$ is compact.
Upon calculation, the eigenvalue vector of \(\mathbf{C}(\theta_0)\) defined in \eqref{eq:c} is \([261.46728326,   5.66471674]\), confirming that Assumption \ref{c_positive} holds.

\subsection{Simulation of Alg.~\ref{alg:fisher}}
\textbf{Settings:} The parameter settings are presented in Table~\ref{tab:parameters}. To validate the active learning for the Stackelberg game detailed in Section~\ref{sec:learning}, we run 50 independent sample paths using Algorithm \ref{alg:fisher}. Each experiment is configured with \( T = 20 \) and \( \lambda = 1 \). As a benchmark, we employ a uniform random strategy, where the leader selects uniformly from \( \mathbf{U}^L \), defined by the box constraints in Table~\ref{tab:parameters}. Algorithm~\ref{alg:fisher} is executed three times, optimizing for D-optimality, A-optimality, and E-optimality, respectively.

\begin{figure*}[htp]
    \centering
    \includegraphics[width=\textwidth]{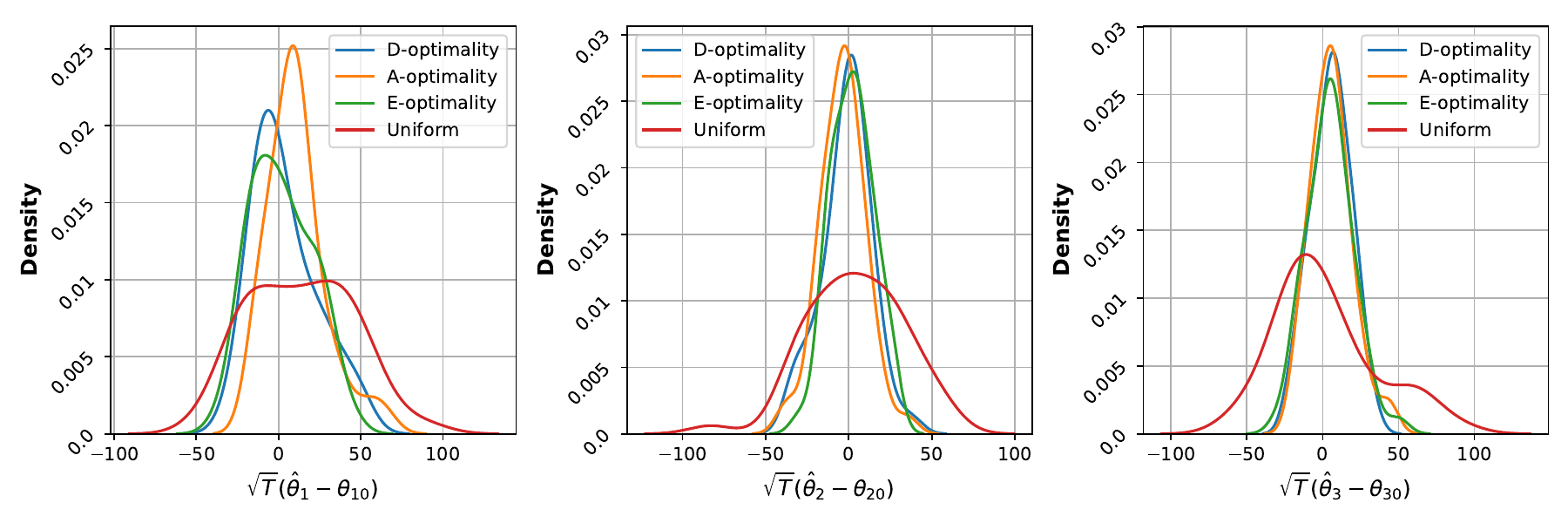}
    \caption{Active learning for Stackelberg games with D-optimality, A-optimality and E-optimality vs. uniform random strategy.} 
    \label{fig:simFisher}
\end{figure*}

\textbf{Results:} The results of active learning for Stackelberg games are depicted in Fig.~\ref{fig:simFisher}, where the horizontal axis represents the regularized estimated bias \( \sqrt{T}(\hat{\mathbf{\theta}}(T) - \mathbf{\theta}_0) \). Each subplot displays the empirical probability density of each element of the estimate derived from samples. It is evident that the estimates obtained through Algorithm \ref{alg:fisher} exhibit a trend towards a normal distribution and consistently converge towards the true parameters. Moreover, the variance of the estimates, whether optimized for D-optimality, E-optimality, or A-optimality, is lower compared to those derived using the uniform random strategy, consistent  with theoretical results.


\begin{table}[ht]
\caption{Parameters settings}
    \label{tab:parameters} 
    \centering
       \begin{tabular}{l|l}
       \hline
        Operational Parameters  & Value   \\
        \hline
        \hline
        $T$ of Alg. 1         & 20 \\
        Num. of sample paths for Alg. 1               & 50 \\
        $T$ of Alg. 2          & 100\\
        Num. of sample paths for Alg. 2   & 300 \\
        $\mathbf{U}^L$               & $\squbra{10,100}\times\squbra{10,100}$ \\
        $\kappa$         & 1e-3 \\
        $\lambda$         & 1 \\
        $\mu_t = \mu_0/t$    & 4e7/t \\
        $\eta$         & 2 \\
        $\alpha$         & 1e3 \\
        \hline
    \end{tabular}
\end{table}

\subsection{Simulation of Alg.~\ref{alg2}}
\textbf{Settings:} To evaluate the active inverse game for the Stackelberg game, focusing on the exploration-exploitation trade-off, we run 300 independent sample paths for Algorithm~\ref{alg2}. Each experiment is conducted with \( T = 100 \) and \( \lambda = 1 \), where the  other parameters are listed in Table~\ref{tab:parameters}. Specifically, Algorithm~\ref{alg2} is implemented exclusively for E-optimality. As a comparison, we test a method without active exploration, which only considers the \textit{exploitation} term in Eq.~\eqref{query2}. After obtaining the estimate \( \hat{\mathbf{\theta}}(t) \), the leader solves
\[
\min_{\mathbf{u}^L\in \mathbf{U}^L} \mathbb{E}_{\mathbf{u}^F\mid\mathbf{u}^L,\hat{\mathbf{\theta}}(t)} \left[J^L(\mathbf{u}^F,\mathbf{u}^L, \hat{\mathbf{\theta}}(t))\right] 
\]
to obtain query action.

\begin{figure}[ht]
    \centering
    \includegraphics[width=\columnwidth]{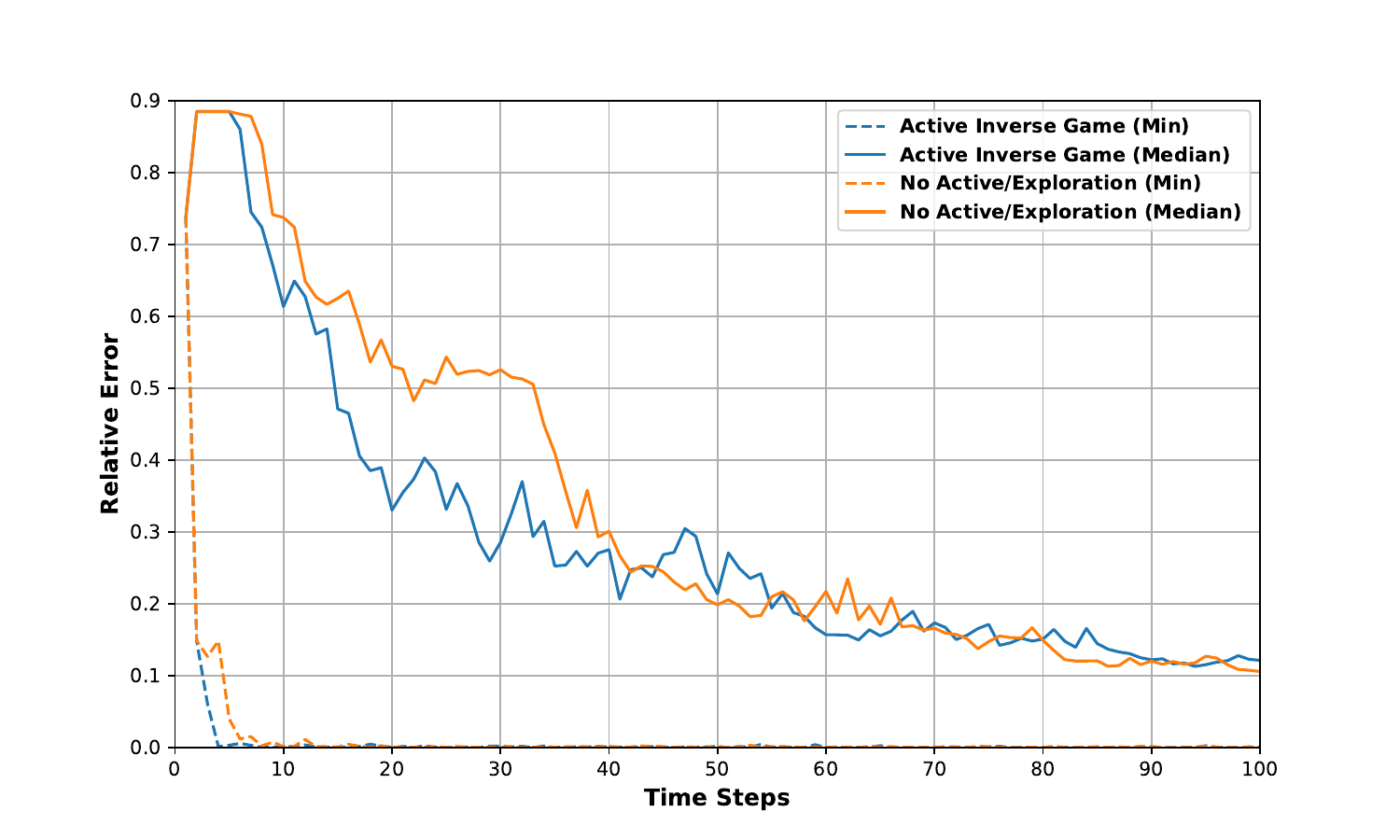}
    \caption{The relative error \( \frac{\|\mathbf{u}^L(T) - \mathbf{u}^{L*}\|}{\|\mathbf{u}^{L*}\|} \) of the active inverse game for Stackelberg games (Algorithm 2) vs. strategy without exploration. The dotted lines represent the best results (Min), and the solid line represents the median of the results (Median)
    of 300 experiments by methods, respectively.}
    \label{fig:simEquil}
\end{figure}

\begin{figure*}[ht]
    \centering
    \includegraphics[width=\textwidth]{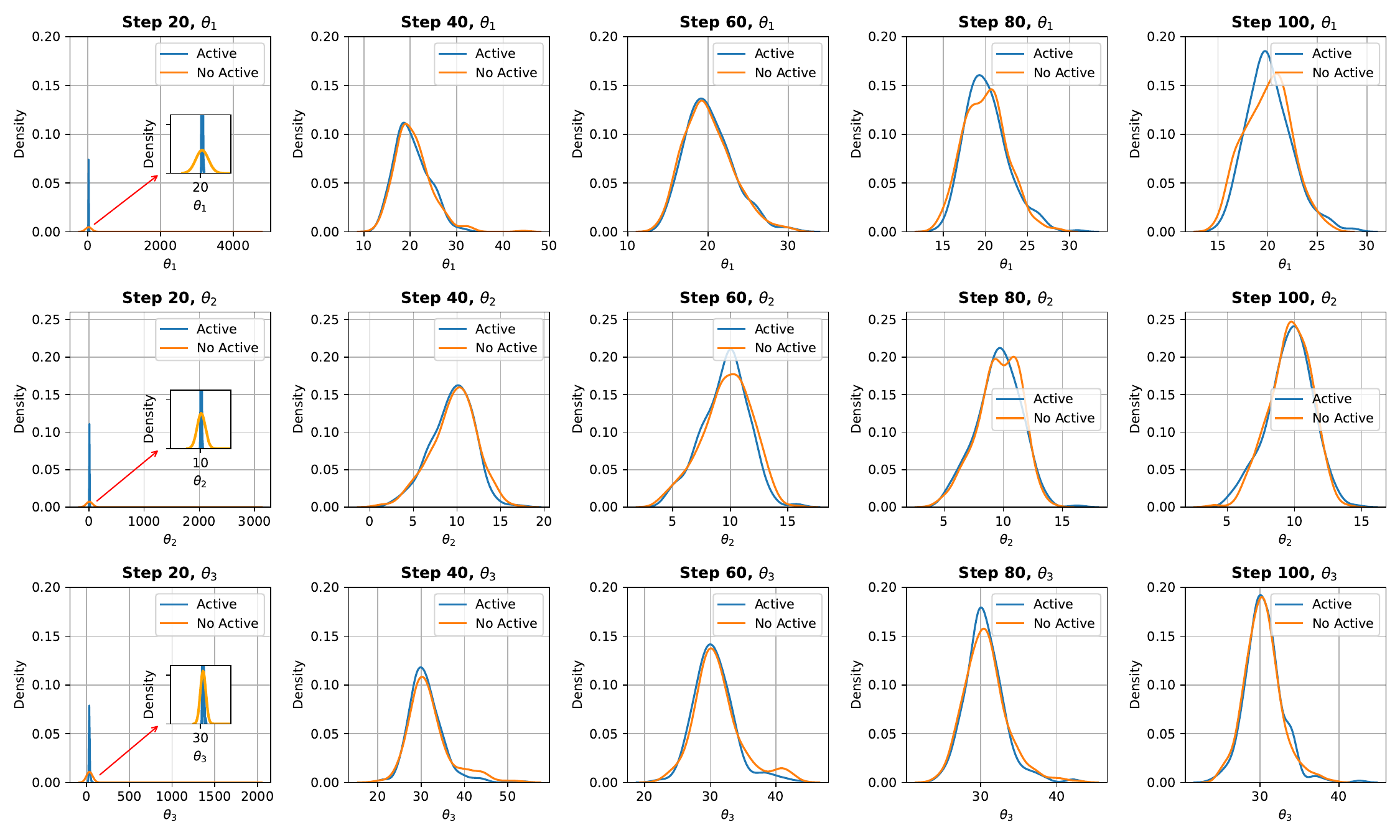}
    \caption{Comparison of parameter estimate by active inverse methods in Stackelberg games and strategy without exploration by minimizing estimated expected cost over time at steps 20, 40, 60, 80 and 100.}
    \label{fig:simEsti}
\end{figure*}

\textbf{Results:} The empirical results are presented in Fig.~\ref{fig:simEquil} and Fig.~\ref{fig:simEsti}. 
Fig.~\ref{fig:simEquil} illustrates the relative error \( \frac{\|\mathbf{u}^L(T) - \mathbf{u}^{L*}\|}{\|\mathbf{u}^{L*}\|} \), where \( \mathbf{u}^L(T) \) represents the leader's action obtained through different methods and \( \mathbf{u}^{L*} \) denotes the true Stackelberg equilibrium defined in \eqref{def:Stackelberg_equilibrium}. It is observed that early-time performance of metrics such as the best, median, and interquartile range favor active inverse game methods over the approach without exploration. However, over time, both methods converge with random oscillations.

Fig.~\ref{fig:simEsti} displays the empirical probability density plots of estimates at different time steps. At time step 20, it is evident that the variance of estimates obtained through the active method is significantly smaller than that of the inactive method, also providing an explanation for the early-stage superiority of the active method. As time progresses, when the variance diminishes sufficiently (indicating \( \rho \approx 0 \), where \( \rho \) is the coefficient of exploration), the exploration component in the query function diminishes, leading to comparable variances for both methods, as depicted in Fig.~\ref{fig:simEsti}. This observation further elucidates why the relative errors of both methods converge in later time steps, as illustrated in Fig.~\ref{fig:simEquil}.

\section{Conclusion}
\label{sec:conclusion}
In this paper, we have developed a framework of active inverse games. We introduced active inverse methods framework specifically in Stackelberg games with bounded rationality, where the leader, acting as a learner, strategically selects actions to better understand the follower's cost functions. We developed an active learning algorithm leveraging Fisher information to maximize information gain about the unknown parameters, ensuring consistency and asymptotic normality with minimal variance in the estimates. Additionally, when considering the leader's cost, we designed an algorithm of active inverse game for Stackelberg games that balances exploration and exploitation, ensuring consistency and asymptotic Stackelberg equilibrium in quadratic games. Finally, we experimentally validated our theoretical findings and advantages of our method through an example of quadratic game.


\appendix

\bibliography{bibfile}
\end{document}